\documentclass[11pt,a4paper]{article}
\usepackage{amssymb,amsfonts,amsmath,amsthm, mathtools}
\usepackage[latin1]{inputenc}
\usepackage[english]{babel}
\usepackage[left=2.7cm,right=2.7cm,bottom=2.67cm,top=2.67cm]{geometry}
\usepackage{dsfont}
\usepackage{mathrsfs}
\usepackage{natbib}
\usepackage{color}
\usepackage{graphicx}
\usepackage{multirow}
\usepackage{multicol}
\usepackage{placeins}
\usepackage[normalem]{ulem}
\usepackage[final]{pdfpages}

\usepackage[implicit=false]{hyperref}

\theoremstyle{plain}
\newtheorem{thm}{Theorem}[section]

\newtheorem{lemma}{Lemma}[section]

\graphicspath{{figs/}}
\theoremstyle{definition}

\newcommand{\E}{\mathds{E}}

\newcommand{\R}{\mathds{R}}
\newcommand{\N}{\mathds{N}}

\newcommand{\bs}[1]{\boldsymbol{#1}}
\renewcommand{\qed}{\hfill \mbox{\raggedright \rule{.07in}{.1in}}}
\renewcommand{\proof}[1]{\noindent\textbf{Proof of Theorem \ref{#1}:}}

\newcommand{\var}{\mathrm{Var}}

\newcommand{\supp}{\mathrm{supp}}
\newcommand{\ase}{\mathrm{ASE}}
\newcommand{\bias}{\mathrm{Bias}}
\DeclareMathOperator*{\argmax}{arg\,max}
\DeclareMathOperator*{\argmin}{arg\,min}
\DeclareMathOperator*{\diag}{diag}
\long\def\sfootnote[#1]#2{\begingroup%
\def\thefootnote{\fnsymbol{footnote}}\footnote[#1]{#2}\endgroup}
\def\bfootnote{\xdef\@thefnmark{}\@footnotetext}

\begin{document}
\pagestyle{myheadings} 
\markboth{Two-steps Production Frontier estimation}{Matsuoka, Pumi, Torrent and Valk}

\thispagestyle{empty}
{\centering
\Large{\bf A two-step approach to production frontier estimation and the Matsuoka's distribution.} \vspace{.5cm}\\
\normalsize{ {\bf
Danilo H. Matsuoka${}^{\mathrm{a,}}$\sfootnote[1]{Corresponding author. This Version: \today},\let\thefootnote\relax\footnote{\hskip-.3cm$\phantom{s}^\mathrm{a}$Mathematics and Statistics Institute and Programa de P\'os-Gradua\c c\~ao em Estat\'istica - Universidade Federal do Rio Grande do Sul.}
Guilherme Pumi${}^\mathrm{a}$, Hudson da Silva Torrent${}^\mathrm{b}$\let\thefootnote\relax\footnote{\hskip-.3cm$\phantom{s}^\mathrm{b}$Mathematics and Statistics Institute - Universidade Federal do Rio Grande do Sul.
}, Marcio Valk${}^\mathrm{a}$
 \\
\let\thefootnote\relax\footnote{E-mails: danilomatsuoka@gmail.com (Matsuoka) guilherme.pumi@ufrgs.br (Pumi); hudsontorrent@gmail.com (Torrent); marcio.valk@ufrgs.br (Valk)}
\let\thefootnote\relax\footnote{ORCIDs:
{0000-0002-9744-8260} (Matsuoka); 
{0000-0002-6256-3170} (Pumi); 
{0000-0002-4760-0404} (Torrent); 
{0000-0002-5218-648X} (Valk)} 
\vskip.3cm
}}
}
\begin{abstract}
In this work, we introduce a deterministic frontier model in which efficiency is governed by the Matsuoka distribution, a parsimonious one-parameter specification on $(0,1)$ designed to reflect patterns typically observed in efficiency data. Based on this formulation, we develop a two-step semiparametric estimation procedure: a nonparametric smoothing for the regression component, followed by a feasible method of moments estimation for the efficiency parameter with plug-in reconstruction of the frontier. Theoretical results establish convergence rates, asymptotic normality, and an oracle property for the parametric estimator of the efficiency parameter. A Monte Carlo study demonstrates that the procedure performs consistently with the theoretical results and improves upon  a fully nonparametric alternative. Applying the method to Brazilian temporary crops with land and agrochemicals as inputs, we find that both regions exhibit isoquants close to the constant elasticity substitution form, but differ in the relative productivity of inputs. Most notably, statistical tests provide evidence that the South is relatively more efficient than the Center-West, highlighting the empirical relevance of the proposed approach.
\vspace{.2cm}\vspace{.2cm}
\end{abstract}
\section{Introduction}
In the economic theory of production, characterizing and estimating efficiency  is essential for performance benchmarking and analyzing productivity. Since the seminal work of \cite{farrell}, a vast literature has developed on production frontier estimation. Two main paradigms are commonly used: the stochastic and the deterministic approaches. In both, the frontier is a fixed function; the stochastic approach treats deviations from the frontier as a mixture of statistical noise and inefficiency, whereas the deterministic approach attributes all deviations to inefficiency. Each paradigm has strengths and limitations \citep[see, e.g.,][]{Bezat_2009,bogetoft2}.

In this work we adopt the \textit{deterministic approach}.
Its appeal in applied work stems from the fact that estimation can proceed under weak assumptions on the production set. Seminal nonparametric estimators in this class include the Data Envelopment Analysis (DEA) of \cite{charnes} and the Free Disposal Hull (FDH) of \cite{deprins}. While simple to implement, these estimators are known to be downward biased and yield piecewise-constant or non-smooth frontiers. To mitigate these issues, \cite{martins} proposed a three-step kernel-based procedure. Subsequent contributions, such as \cite{fang} and \cite{chen}, incorporated robustness to outliers and shape constraints.

We depart from this literature by introducing a deterministic frontier model in which the efficiency term follows the \emph{Matsuoka distribution}, a one-parameter law supported on $(0,1)$ that flexibly accommodates features commonly observed in frontier applications (e.g., skewness and heavy left tails). While it is a particular case of the log-gamma family of \cite{consul}, the use of the Matsuoka distribution in production frontiers is, to our knowledge, novel. We develop the distributional groundwork needed for estimation, including basic properties and additional results useful in practice.

Building on this structure, we propose a \emph{two-step semiparametric estimation} method inspired by multiplicative specifications in the spirit of Cobb-Douglas functions. In the first step, the regression function is estimated nonparametrically. In the second step, the efficiency parameter is estimated by a feasible method of moments based on residuals, and the production frontier is recovered by a plug-in transformation.

\medskip
\noindent\textbf{Contributions.}
This paper makes several contributions to the production frontier literature. First, we introduce a new deterministic frontier framework in which efficiency follows the Matsuoka distribution, a parsimonious yet flexible law on $(0,1)$ that was designed to capture empirical regularities of efficiency data, such as concentration near the frontier and the absence of perfectly efficient units. Second, we propose a practical two-step semiparametric estimation method that makes use of the parametric specification for efficiency. Third, we develop a rigorous asymptotic theory for the proposed estimator, covering both univariate and bivariate input models, and establishing convergence rates and asymptotic normality for all components. A remarkable result is that the second-step parametric estimator satisfies an \emph{oracle property}, meaning that its large-sample behavior is unaffected by the preliminary nonparametric estimation. We also conduct a comprehensive Monte Carlo study to document the finite-sample performance of the method and to compare it with fully nonparametric alternatives. 
Finally, we provide an empirical application to Brazilian temporary crops, analyzing production with two inputs - \emph{land} and \emph{agrochemicals} - for the South and Center-West regions. The estimated isoquants are close to the constant elasticity of substitution (CES) form in both cases, yet they reveal distinct critical inputs across regions, offering policy-relevant insights.

The paper is organized as follows. Section \ref{pf} presents the model. Section \ref{seca} introduces the proposed two-step estimation method. Section \ref{at} develops the asymptotic theory. Section \ref{sim} reports a Monte Carlo simulation study to investigate the finite-sample properties of the estimator. Section \ref{ea} presents the empirical application. Section \ref{conc} concludes.
%
\section{Production Frontier Model}\label{pf}
In economic theory, a production function is defined as the maximum output that can be obtained from a set of inputs, given the technology available to the firms. Formally, let $\bs x\in\R_+^m$ be a vector of inputs used to produce output $y\in\R_+$, and define the production set  as $\Psi=\{(y,\bs x)\in \R_+^{m+1}: \bs x \text{ can produce } y \}$, where $\R_+=\{z\in\R:z\geq 0\}$. For all $\bs x\in\R_+^m$, the production function (or frontier) associated with $\Psi$ is defined as $f(\bs x)=\sup\{y\geq 0: (y,\bs x)\in\Psi\}$. Moreover, for all pairs $(y,\bs x)\in\Psi$, let the efficiency be measured by the ratio $y/f(\bs x)$.

From a statistical viewpoint, suppose that a random sample of production units $(Y_i,\bs{X}_i)_{i=1}^n$  is observed, where each pair takes values in  $\Psi$ and follows the model
\begin{equation}\label{equa1}
    Y_i=f(\bs{X}_i)R_i, \quad i\in\{1,\cdots,n\},
\end{equation}
where $R_i$ is a random variable taking values in $(0,1)$ whose distribution will be discussed below, and  $f:\R^m\to\R$ is a function that is multiplicative on unknown nonnegative smooth components $f_j$, i.e.,
\begin{equation}\label{equacob}
    f(\bs{X}_i)=\prod_{j=1}^m f_j(X_{i,j}),
\end{equation}
where $X_{i,j}$ denotes the $j$th component of $\bs X_i$. In this context, $R_i$ is the efficiency, $\bs{X}_i$ is the input vector, $Y_i$ is the output and $f$ is the production frontier. When  $R_i$ is close to $1$, the observed output $Y_i$ lies close to the frontier $f(\bs{X}_i)$, indicating a highly efficient firm. Conversely, when $Y_i$ is far from the frontier $f(\bs{X}_i)$, $R_i$ is small, indicating low efficiency. For simplicity, we assume that $R_i$ is independent of $\bs{X}_i$ and that $P(\bs X_i\in B)=1$, for some compact set $B\coloneqq\prod_{j=1}^m[a_j,b_j]\subset\R^m$.

The functional form in \eqref{equacob} includes the well-known Cobb-Douglas production function where $f$ takes the form
$f(\bs{X}_i)=a_0\prod_{j=1}^m X_{i,j}^{a_j}, \ i\in\{1,\cdots,n\}$,
with parameters $a_0, \dots, a_m$ quantifying the responsiveness of the output to changes in each input. This framework is useful when modeling interactions among input variables. A key feature of this model is its log-additive structure, which alleviates the curse of dimensionality in multivariate nonparametric estimation \citep[see][]{stone}.  Taking logarithms yields
\begin{equation}\label{equa2}
      Z_i=g(\bs{X}_i) +\epsilon_i, \quad i\in\{1,\cdots, n\},
\end{equation}
where $Z_i\coloneqq -\ln(Y_i)$ is the log-transformed data,  $\epsilon_i\coloneqq -\ln(R_i) -\E(-\ln(R_i))$ is a zero mean error term, and the regression function becomes
\begin{equation}\label{equa2__}
    g(\bs{X}_i)= g_0+\sum_{j=1}^{m}g_j(X_{i,j})
\end{equation}
with  $g_0=\E(-\ln(R_i))$ and $g_j(X_{i,j})\coloneqq -\ln(f_j(X_{i,j}))$, $j\in\{1,\cdots,m\}$.  For $m>1$, we impose the standard identifiability constraints $\E\{g_j(X_{i,j})\}=0$ for all $j$, so that $\E(Z_i)=g_0$. Accordingly, we work with centered versions of $g_1,\dots,g_m$ with $g_0$ capturing the mean level.

Although the model in \eqref{equa1} can, in principle, be estimated without specifying a probabilistic law for $R_i$, in this study we postulate a particular distribution for efficiency. This choice not only improves the tractability of the estimation problem, but also facilitates the interpretability of technical efficiency, since the distributional shape can be directly linked to economic notions of firms operating close to or far from the frontier. The next section introduces the distributional assumptions that underlie our framework.
\subsection{Probability Distribution of the Efficiency}
A general stochastic frontier \citep{greene2003} related to model \eqref{equa2} can be written  as
\begin{equation}\label{equadet}
    Z_i=g(\bs X_i)+v_i+\epsilon_i
\end{equation}
where $v_i$ is a symmetric zero mean random variable capturing random fluctuations in the production process that are independent of technical inefficiency $-\ln(R_i)$, a one-sided disturbance with positive mean. Setting $v_i=0$ yields the \textit{deterministic frontier model}, which is the focus of this work. Classic choices for technical inefficiency include half-normal and  exponential distributions \citep{aigner,meeusen}, and the Gamma distribution \citep{stevenson1980,greene1990}, the latter offering a richer and more flexible parametrization.

We start modeling the inefficiency by assuming that the random variable  $-\ln(R_i)$ follows a Gamma distribution with shape and rate parameters $\alpha>0$ and $p>0$, respectively. This assumption allows for the interpretation that technical inefficiency is a sum of independent sources of inefficiency, provided that each of them is Gamma distributed with common rate parameter. An important consequence is that the efficiency, $R_i$, follows the Log-Gamma distribution introduced by \cite{consul}, whose density function is given by
 \begin{equation}\label{eqloggam}
 	f_{\alpha,p}(x)=\frac{p^{\alpha}}{\Gamma(\alpha)} x^{p-1}(-\ln(x))^{\alpha-1} I(0<x<1), \quad p,\alpha>0.
 \end{equation}
Firms in competitive markets often have technical efficiency close to one, a pattern particularly evident among surviving firms in many datasets as shown by \cite{berger1997}. Our modeling approach is specifically designed to reflect this empirical reality. To keep the model simple and aligned with empirical settings where technical efficiency is concentrated near one, we will fix a value for $\alpha$.
This choice is guided by two criteria: (i) it is an unattainable ideal for a firm to have full efficiency, with only a few operating near the frontier; and (ii) firms should not be overly concentrated far from the frontier.

The key components of the density \eqref{eqloggam} are the power $x^{p-1}$ and the logarithmic term $(-\ln(x))^{\alpha-1}$. Note that the criteria (i) can be translated as requiring that $f_{\alpha,p}(x)\to 0$ as $x$ approaches  1 from the left, which is satisfied if, and only if, $\alpha>1$. 
Moreover, as the parameter $\alpha$ increases, the logarithmic term $(-\ln(x))^{\alpha-1}$ dominates the power term $x^{p-1}$ when $x$ is close to zero,  leading to a heavier left tail for any $p>0$. Therefore, the criteria (ii) argues that $\alpha$ cannot be taken too big. In summary, we have to choose a value $\alpha>1$ that is not too far from one. The value $\alpha=1.5$ balances these criteria. For this value, the logarithm term becomes $\sqrt{-\ln(x)}$ and $\Gamma(1.5)=\sqrt{\pi}$. The induced efficiency density is
\begin{equation}\label{eq1}
f_p(x)\coloneqq  2\sqrt{\frac{-p^3 \ln(x)}{\pi}} x^{p-1} I(x\in(0,1)),
\quad x\in\R,
\end{equation}
denoted $R_i\sim M(p)$ and referred to as the \emph{Matsuoka distribution} with parameter $p>0$.

A more detailed treatment of Matsuoka's distribution is provided in Appendix B. A direct implication of the assumption $R_i \sim M(p)$ is that $\E(-\ln(R_i))=3/(2p)$, which establishes the exact log-linear form used in \eqref{equa2} and \eqref{equa2__}. We highlight that when  $p$ is greater than approximately 1.72, the distribution becomes left-skewed, allowing for the modeling of left-skewed efficiency with high resolution in $p$. The paper accompanies a supplementary material where further properties of the Matsuoka's distribution are derived.

In short, by modeling efficiency with the Matsuoka distribution, characterized by an unknown parameter $p>0$, we obtain a flexible yet tractable specification that aligns with observed efficiency patterns in production data. This formulation reduces the estimation problem to two main components: recovering the nonparametric function $g$ and estimating the distributional parameter $p$. In the following section, we outline a two-step procedure that addresses both tasks within a unified framework.
\section{Proposed two-step estimation}\label{seca}
\subsection{Step 1: estimation of the regression function $g$}\label{sec.a1}
The first step consists of applying a nonparametric estimator  with dependent variable $Z_i$ and regressors $\bs{X}_i$, $i\in\{1,\cdots,n\}$. Any reasonable nonparametric method could be used at this stage. In this study, we restrict our attention to the local linear smoother \citep[see][for a comprehensive overview of local polynomial regression]{wand_jones, tsybakov}, for the univariate-input setting, and the classical and smooth backfitting schemes, hereinafter referred to as CBS \citep{buja,hastie} and SBS \citep{mammen} respectively, for bivariate inputs.

Consider the $n$-vector $g_j(\bs x_j)\coloneqq ( g_j(x_{1,j}),\cdots, g_j(x_{n,j}))^\intercal$  evaluated at grid points $\bs x_j=(x_{1,j},\cdots,x_{n,j})^\intercal$, $j\in\{1,\cdots,m\}$,  and let $\bs{Z}\coloneqq (Z_1,\cdots,Z_n)^\intercal$ with $Z_i$ being given by equation \eqref{equa2} for all $i\in\{1,\cdots,n\}$.

We estimate each vector $g_j(\bs x_j)$ using a linear smoother of the form
\begin{equation}\label{equa3_}
 \hat g_j(\bs x_j)\coloneqq W_j  \bs{Z}^\ast , \quad j\in\{1,\cdots,m\},
 \end{equation}
where $W_j$ is an $n\times n$ smoothing matrix, $Z^\ast\coloneqq \bs{Z} - \bs{\bar Z}$,  with $\bs{\bar Z}\coloneqq \bar Z I(m>1)\bs 1_n$, for $\bs 1_n\coloneqq (1,\cdots,1)^\intercal\in\R^n$ and $\bar Z=\frac1n\sum_{i=1}^n Z_i$. Note that when $m>1$, the intercept $g_0$ is implicitly estimated using the sample mean. From  \eqref{equa3_}, an estimator for the regression function $g$ is
\begin{equation*}
    \hat g(x_{i,1},\cdots,x_{i,m})=\bs e_{i,n}^\intercal\sum_{j=1}^m\hat g_j(\bs x_j)+\bar ZI(m>1),\quad \forall i\in\{1,\cdots,n\},
\end{equation*}
where $\bs e_{i,n}$ is the $n$-vector with $1$ in the $i$th coordinate and $0$ everywhere else.

We now introduce notation used in the definition of the proposed estimator. Let $K$  be a kernel function and, for bandwidths $h_1,\cdots,h_m$, let
\begin{equation}\label{oka}
K_{h_j}(u)\coloneqq \frac1{h_j}K\bigg(\frac{u}{h_j}\bigg).
\end{equation}
For $z\in\R$, let
\begin{equation*}
    A_{z}^j\coloneqq \left[
	\begin{array}{cc}
		1 & (X_{1,j}-z)  \\ 
		\vdots & \vdots  \\ 
		1  &  (X_{n,j}-z)
	\end{array}
	\right],
\end{equation*}
$D_z\coloneqq \diag(K_{h_j}(X_{1,j}-z),\cdots,K_{h_j}(X_{n,j}-z))$, and let $\bs w_{j,z}$ be the equivalent kernel for $j$th covariate at the point $z$ defined by $\bs w_{j,z}^\intercal \coloneqq \bs e_{1,2}^\intercal({A_z^j}^\intercal D_zA_z^j)^{-1}{A_z^j}^\intercal D_z$, provided that $({A_z^j}^\intercal D_zA_z^j)^{-1}$ exists. Finally, let
\begin{equation}\label{equa3}
    S_j=(\bs w_{j,x_{1,j}},\cdots,\bs w_{j,x_{n,j}})^\intercal, \quad j\in\{1,\cdots,m\}.
\end{equation}
\subsubsection*{Local Linear}
For $m=1$, we estimate the regression function $g(\cdot)=g_0-\ln(f(\cdot))$ through the local linear estimator implying that the smoothing matrix is given by $W_1=S_1$ as defined in equation \eqref{equa3} for $j=1$.
 Thus, we estimate $g$ at point $x_{i,1}$ by
\begin{equation}\label{equa13_}
    \hat g (x_{i,1})=\bs e_{i,n}^\intercal S_1\bs{Z}, \quad i\in\{1,\cdots,n\}.
\end{equation}
\subsubsection*{Classical Backfitting}
For $m=2$, the regression function  $ g(x_{i,1},x_{i,2})=g_0+g_1(x_{i,1})+g_2(x_{i,2})$ is estimated using  CBS with local linear smoothers. To ensure identification and estimation, the model given by \eqref{equa2}--\eqref{equa2__} must satisfy $\E(Z_i)=g_0$ and $\E(g_j(X_{i,j}))=0$, for all $j\in\{1,\cdots,m\}$ and $i\in\{1,\cdots,n\}$.  Therefore, we need to rewrite equation \eqref{equa2__} with $g_0=\frac3{2p}-\E\big(\sum_{j=1}^{m}\ln\big(f_j(X_{i,j})\big)\big)$ and $g_j(X_{i,j})=\E\big(\ln\big(f_j(X_{i,j})\big)\big)-\ln\big(f_j(X_{i,j})\big)$, $j\in\{1,\cdots,m\}$. As shown in \cite{opsomer} and \cite{hastie}, the related smoothing matrices admit explicit expressions and are given by
\begin{align*}
     W_1&=I_n-(I_n-S_1^\ast S_2^\ast )^{-1}(I_n-S_1^\ast );\\
     W_2&=I_n-(I_n-S_2^\ast S_1^\ast )^{-1}(I_n-S_2^\ast ),
\end{align*}
 provided that the inverses exist, where $S_j^\ast \coloneqq S_j - \big(\frac1n\sum_{i=1}^n S_j\big)\bs 1_n$,  with $S_j$ being defined analogously as in \eqref{equa3}. As explained by  \cite{opsomer}, the backfitting estimators for the functions $g_j$ at the observation points are obtained nonparametrically by solving a system of normal equations. Corollary 4.3 of \cite{buja} shows that $\| S_1^\ast S_2^\ast \|< 1$ is a sufficient condition for the existence of unique backfitting estimators when $m=2$, for any matrix norm $\|\cdot\|$.\footnote{The conditions for existence and uniqueness of CBS when $m>2$ are provided in Lemma 2.1 of \cite{opsomer}. In practice, the $m$-variate CBS can be approximated by the backfitting algorithm \citep[ Chapter 11]{wilcox}.} Therefore, we can estimate the regression function $g$ at $(x_{i,1},x_{i,2})$ by
   \begin{equation*}
       \hat g(x_{i,1},x_{i,2})=\bar Z+\bs e_{i,n}^\intercal(W_1+W_2)\bs{Z}^\ast , \quad i\in\{1,\cdots,n\}.
   \end{equation*}
\subsubsection*{Smooth Backfitting}
Still, for $m=2$, the additive regression function $g(x_{i,1},x_{i,2})$ can be alternatively estimated using the smooth backfitting estimator (SBS) based on Nadaraya-Watson smoothers. Denote the support of each covariate $X_j$  by  $B_j$,  and the support of $\bs X$ by  $B=B_1\times B_2$. In addition, let $\bs{x}=(x_1,x_2)$ and $\bs h=(h_1,h_2)$. Considering \eqref{oka}, let
\begin{equation*}
K_{h_j}(x_j,u_j)\coloneqq\frac{K_{h_j}(x_j-u_j)}{\int_{B_j}K_{h_j}(x_j-w_j)dw_j},\quad \mbox{for }\ j\in\{1,2\},
\end{equation*}
which implies that $\int_{B_j}K_{h_j}(x_j,u_j)du_j=1$, for all $x_j\in B_j$, for $j\in\{1,2\}$. Define a two-dimensional product kernel by setting $K_{\bs h}(\bs x,\bs X_{i}) \coloneqq K_{h_1} (x_1,X_{i,1}) K_{h_2}(x_2,X_{i,2})$  and let $\hat f_{\bs{X}}(\bs x)=n^{-1}\sum_{i=1}^n  K_{ \bs h}(\bs x,\bs X_{i})$ denote Nadaraya-Watson's joint kernel estimator with marginals $\hat f_{X_j}(x_j)= n^{-1}\sum_{i=1}^n K_{h_j}(x_j,X_{i,j})$, $j\in\{1,2\}$.

Let $\mathscr{A}_B$ denote the space of all additively composed real functions $(g_0,g_1,g_2)$ of the form $g(\bs x)=g_0+ g_1(x_1)+ g_2(x_2)$, with $\bs x=(x_1,x_2)\in B$ and $g_0\in\R$, and such that
\begin{equation*}
 \int_{B_j}  g_j(x_j)\hat f_{X_j}(x_j)dx_j=0, \quad \mbox{for }\ j\in\{1,2\}.
\end{equation*}
The SBS estimator is defined as
 \begin{equation}\label{eqsb16}
     (\hat g_0,\hat g_1,\hat g_2) \coloneqq\argmin_{(g_0,g_1,g_2)\in \mathscr{A}_B}\bigg\{\int_{B}\sum_{i=1}^n \big( Z_i-g_0- g_1(x_1)- g_2(x_2)\big)^2 K_{\bs h}(\bs{x},\bs{X}_{i})d\bs{x}\bigg\}.
 \end{equation}
Viewing \eqref{eqsb16} as the projection of the Nadaraya-Watson regression smoother
\begin{equation*}
 g(\bs{x})=\frac{\sum_{i=1}^n K_{\bs h}(\bs x,\bs X_{i})Z_i}{\sum_{i=1}^n K_{\bs h}(\bs x,\bs X_{i})}
\end{equation*}
onto the subspace of additive functions $\{ g\in L_2(\hat f_{\bs{X}} ): g(\bs{x})=g_0+g_1(x_1)+ g_2(x_2)\}$, \cite{mammen} characterize the solution of \eqref{eqsb16} by the following system:
\begin{align}\label{eqsb17}
\begin{cases}
   \displaystyle{\hat g_j(x_j)=\tilde g_j(x_j)-\sum_{\substack{k=1 \\ k\neq j}}^m\int_{B_k} \hat g_k(x_k)\frac{\hat f_{\bs{X}}(x_j,x_k)}{\hat f_{X_j}(x_j)}dx_k-\hat g_{0},}\\[.3cm]
    \displaystyle{\int_{B_j} \hat g_j(x_j) \hat f_{X_j}(x_j)dx_j=0},
\end{cases}
\end{align}
for $j\in\{1,2\}$, where
\begin{equation}
\tilde g_j(x_j)=\frac{\sum_{i=1}^n K_h(x_j,X_{i,j})Z_i}{n\hat f_{X_j}(x_j) },
\end{equation}
and $\hat g_0$ is taken as the sample mean $\bar Z$, which is a $\sqrt{n}$-consistent estimate of the population mean.

In practice, the solutions of \eqref{eqsb17} can be approximated using an iterative backfitting algorithm \citep[for further details and a comparison with CBS, see][]{nielsen}. Convergence and uniqueness for this algorithm are established by Theorem 1 of \cite{mammen}.  Once the solution of \eqref{eqsb17} is well approximated by $\{\bar Z, \hat g_1(x_{i,1}),\hat g_{2}(x_{i,2})\}$ via backfitting algorithm, we can estimate $g$ at $(x_{i,1},x_{i,2})$ by
\begin{equation*}
\hat g(x_{i,1},x_{i,2})=\bar Z+\hat g_1(x_{i,1})+\hat g_2(x_{i,2}), \qquad \mbox{for }\ i\in\{1,\cdots,n\}.
\end{equation*}
\subsection{Step 2: estimation of the parameter $p$ and plug-in frontier}\label{sec.a2}
When the regression function $g$ in \eqref{equa2} is known, we have that  $Z_i-g(\bs{X}_i)=\epsilon_i$.
We then construct a method of moments estimator by equating the second sample moment with $\E(\epsilon_i^2)=3/(2p^2)$, which leads to
\begin{equation*}
    \tilde p=\sqrt{\frac{3n}{2\sum_{i=1}^n\epsilon_i^2}}.
\end{equation*}
Clearly, $\tilde p$ is not feasible since $\{\epsilon_i\}_{i=1}^n$ is unobserved. We approximate each $\epsilon_i=Z_i-g(\bs{X}_i)$
 by replacing $g$ with the first-step estimate $\hat g$, which in turn results in $\hat \epsilon_i= Z_i-\hat g(\bs{X}_i),\ i\in\{1,\cdots,n\}$. Therefore, a feasible estimator for $p$ is
\begin{equation}\label{equa5}
    \hat p=\sqrt{\frac{3n}{2\sum_{i=1}^n\hat \epsilon_i^2}}.
\end{equation}
\paragraph{Plug-in frontier reconstruction.}
By definition, $g(x)=-\ln(f(x))+3/(2p)$, which is equivalent to
\begin{equation}\label{equa6}
    f(x)=\exp\bigg\{\frac{3}{2p}-g(x)\bigg\}.
\end{equation}
Given $\hat g$ and $\hat p$, the production frontier is recovered by the plug-in rule
\begin{equation}\label{eq11}
    \hat f(x)=\exp\!\left\{\frac{3}{2\hat p}-\hat g(x)\right\}.
\end{equation}
As this is a deterministic transformation of $(\hat g,\hat p)$, we treat it as part of the second step rather than a separate estimation stage.\footnote{This follows the usual convention in the nonparametric frontier literature \citep[see][]{martins,martins2013,chen}.}

\subsection{Asymptotic theory}\label{at}
In this section, we  derive the asymptotic properties of the estimators presented in Sections \ref{sec.a1}--\ref{sec.a2}. We  focus mainly on the univariate local linear estimator and the multivariate smooth backfitting estimator, for which most of the results required for our theory are available. Throughout this study, we use $C$ to denote a generic positive constant that may have different values in different occurrences, and the symbol $\overset{a}{\approx}$ to denote asymptotic equivalence.

Consider the case of univariate inputs, $X\in\R_+$, and denote the density of $X$ by $f_X$. We make the following assumptions:
\begin{itemize}
    \item[\textbf{A1.}] The random variable $X$ has compact support $[a,b]$, for some $b>a$, and its density $f_X$ satisfies
    \begin{equation*}
    0<\min_{x\in [a,b]} \big\{f_X(x)\big\}\leq\sup_{x\in [a,b]}\big\{f_X(x)\big\}\leq C;
    \end{equation*}
    \item[\textbf{A2.}]   The kernel $K: \R\to\R$ is symmetric, satisfies $\displaystyle{\sup_{u\in\R}}\big\{\big|u^iK(u)\big|\big\}\leq C$ for $i\in\{0,1,2\}$, and $\int_{\R}\big|u^jK(u)\big|du\leq C$ for $j\in\{0,1,2,4\}$. In addition, for some $\kappa<\infty$ and $L<\infty$, either
 \begin{itemize}
  \item[(i)] $K(u)=0$ for all $|u|>L$ and $\big|K(u)-K(u')\big|\leq \kappa|u-u'|$ for all $u,u'\in\R$;
  \end{itemize}
or
\begin{itemize}
 \item[(ii)] $K$ is differentiable, $\big|\partial K(u)/\partial u\big|\leq \kappa$, and $\big|\partial K(u)/\partial u\big|\leq \kappa|u|^{-v}$, for all $|u|>L$ and some $v>1$.
 \end{itemize}
	\item[\textbf{A3.}]  There exists $s>2$ such that $\E(|Z|^s)<C$ and
 $\displaystyle{\sup_{x\geq0}}\big\{\E(|Z|^s|X=x)f_X(x)\big\}\leq C$;
 \item[\textbf{A4.}]  The second derivatives of $f_X(x)$ and $f_X(x)g(x)$ exist and are uniformly continuous and bounded;
	\item[\textbf{A5.}]   The bandwidth satisfies $h=o(1)$ and $\ln(n)/(nh)=o(1)$, as $n\to\infty$.
\end{itemize}
Assumption \textbf{A1} requires that $X$ is supported on some interval $[a,b]$ and has density which is bounded above by some constant $C$ and is bounded away from zero.  \textbf{A2} requires that function $K$ is bounded and integrable, and is either Lipschitz with compact support, or has a bounded derivative with an integrable tail. Therefore, most of the commonly used kernels are allowed, including the Epanechnikov kernel (or more generally, polynomial kernels of the form $c(p)(1-x^2)^p$), the Gaussian kernel, and some higher order kernels \citep[e.g., those in ][]{muller,wand90}. \textbf{A3} requires uniform moment bounds for $Z$, and controls the tail behavior of the conditional expectation $\E(|Z|^s|X=x)$, which is allowed to diverge as $x\to\infty$, but at a rate no faster than $1/f_X(x)$. The uniform convergence results of \cite{masry} assume that the second derivatives of $g(x)$ are uniformly bounded. Instead, \textbf{A4} assumes that the second derivatives of the product $f_X(x)g(x)$ are uniformly bounded which is less restrictive for the local linear estimation, as pointed out by \cite{hansen}.  Assumption \textbf{A5} is a strengthening of the usual condition that $h\to0$ and $nh\to\infty$ as $n\to\infty$.

In the following, we give the convergence results for some of the nonparametric estimators of $g$ introduced in Section \ref{sec.a1}. Theorem \ref{teo1} provides the rate of uniform convergence in probability and establishes the asymptotic normality for the local linear estimator defined in \eqref{equa13_}.  Mathematical proofs are deferred to Appendix A.
\begin{thm}\label{teo1}
Suppose that \textbf{A1}- \textbf{A5} hold. Then
\begin{equation}\label{equa24}
\max_{x\in[a,b]}\big\{\big|{\hat g(x)-g(x)}\big|\big\}=O_p\bigg(h^2+\sqrt{\frac{\ln(n)}{nh}}\bigg).
\end{equation}
If in addition the second derivative $g''(\cdot)$ is a continuous function and $nh^3\to\infty$, then for all points $x$ in the interior of $\supp(f_X)$ for which the function $H(y)\coloneqq \E(Z^4|X=y)$ is bounded on a neighborhood of $x$,  we have that
\begin{equation}\label{equa24_}
\sqrt{nh}\big[\hat g(x)-g(x)-B(x)\big]\overset{d}{\longrightarrow}N(0,V(x)),
\end{equation}
where
\[B(x)\coloneqq \frac{g''(x)h^2}2\int_\R u^2 K(u)du\mbox{ , } \quad V(x)\coloneqq \frac{\sigma^2_\epsilon}{f_X(x)} \int_\R K(u)^2du, \quad \mbox{and } \quad\sigma^2_\epsilon\coloneqq\var(\epsilon_1)=\frac{3}{2p^2}.\]
\end{thm}
Methods for approximating the asymptotic conditional bias and variance in \eqref{equa24_} are discussed in Sections 4.4 and 4.5 of \cite{fan_gijbels}. The bias term $h^2$ in \eqref{equa24} can be removed by employing a second-order kernel. As a result, the convergence achieves the minimax-optimal rate established by \cite{stone2}.  The uniform convergence in Theorem \ref{teo1} is established on a compact interval $[a,b]$, for simplicity. However, one may consider a design density with unbounded support $f_X$, extending uniformity to growing intervals that expand slowly with $n$. This incurs a penalty in the convergence rate \citep[see Theorem 10 of][]{hansen}. The next theorem gives the convergence rate for estimator $\hat p$ introduced in \eqref{equa5} of Section \ref{sec.a2}.
\begin{thm}\label{teo2}
Suppose that \textbf{A1}-\textbf{A5} hold. In addition, assume that the second derivative $g''(\cdot)$ is a continuous function, that the kernel function is of type \textbf{A2}(i) and that the bandwidth satisfies $h\overset{a}{\approx} c_\eta n^{-\eta}$, $c_\eta>0$, for $\eta\in[1/8+\delta,1/3-\delta]$, for some arbitrarily small $\delta>0$.  Then
\begin{equation*}
\hat p -p=O_p\bigg(\frac{1}{\sqrt{n}}\bigg), \quad
\mbox{ and }\quad
\sqrt{n}(\hat p-p)\overset{d}{\longrightarrow} N(0,3p^2/2).
\end{equation*}
\end{thm}
Intuition may suggest that the unfeasible estimator $\tilde p$ should be asymptotically more accurate than the feasible estimator $\hat p$. The rationale is that $\hat p$ is calculated from a preprocessed data, where the regression function $g$ has been approximately removed, potentially introducing   additional noise in the estimation of $p$. However, Theorem \ref{teo2} shows that this intuition is misguided. A closer look at its proof reveals that the noise from the first step plays a role in the estimation of $p$  through the term $T_{1,n}=\sum_{i=1}^n(\hat \epsilon_i^2-\epsilon_i^2)/n$. This term is shown to be of order $o_p(1/\sqrt{n})$ for a wide range of bandwidth rates, including the optimal rate $n^{-1/5}$. Since this component is asymptotically negligible, only the unfeasible estimator $\tilde p$   contributes to the limiting behavior. In addition, as $\sqrt{n}(\tilde p-p)\overset{d}{\longrightarrow} N(0,3p^2/2)$, Theorem \ref{teo2} implies that $\hat p$ has the oracle property, that is, it converges to the same asymptotic distribution as $\tilde p$.
%
%
\begin{thm}\label{teo2.n}
Suppose that \textbf{A1}-\textbf{A5} and the conditions of  Theorems \ref{teo1} and \ref{teo2} hold. Then
\begin{equation*}
\max_{x\in[a,b]}\big\{\big|\hat f(x)-f(x)\big|\big\}=O_p\bigg(h^2+\sqrt{\frac{\ln(n)}{nh}}\bigg).
\end{equation*}
 In addition, assume that the bandwidth satisfies $h\overset{a}{\approx} c_\eta n^{-\eta}$ for $\eta\in[1/8+\delta,1/5]$ for some arbitrarily small $\delta>0$. Then, for all  $x$ in the interior of $\supp(f_X)$ for which the function $H(y)\coloneqq \E(Z^4|X=y)$ is bounded on a neighborhood of $x$,  we have that
\begin{align*}
    \sqrt{nh}\big(\hat f(x)-f(x)\big)\overset{d}{\longrightarrow}f(x)N\big(\mu_x,V(x)\big).
\end{align*}
 where $\displaystyle{\mu_x:=\bigg(\frac{c_{\eta}^{1/2}g''(x)}2\displaystyle\int_\R u^2 K(u)du\bigg)I(\eta=1/5).}$
\end{thm}
\noindent Because the estimators $\hat p$ and $\hat g$ are directly plugged into the expression for $\hat f(x)$, the asymptotic behavior is naturally governed by the order of the nonparametric component -- a fact established in Theorem \ref{teo2.n}.

Considering  the case of multivariate inputs, $\bs{X}\in\R_+^m$, denote the density of $\bs{X}$ by $f_{\bs{X}}$. We make the following set of assumptions:
\begin{itemize}
    \item[\textbf{B1.}] The $m$-dimensional random vector $\bs{X}$ has compact support $[0,1]^m$, and its density $f_{\bs{X}}$ satisfies
    \begin{equation*}
    0<\min_{\bs{x}\in [0,1]^m} \big\{f_{\bs{X}}(\bs{x})\big\}\leq \max_{\bs{x}\in [0,1]^m} \big\{f_{\bs{X}}(\bs{x})\big\}\leq C;
    \end{equation*}
    \item[\textbf{B2.}]   The kernel $K: \R\to\R$ is symmetric and $K(u)=0$ for all $|u|>1$. In addition, $K$ is Lipschitz continuous, that is, for some $\kappa_2>0$ and all $u,u'\in\R$, $|K(u)-K(u')|\leq \kappa_2|u-u'|$;
	\item[\textbf{B3.}]  There exists $s_2>5/2$ such that $\E(|Z|^{s_2})<\infty$;
 \item[\textbf{B4.}]  The second partial  derivatives of $g(\bs{x})$ and the first partial derivatives of $f_{\bs{X}}(\bs{x})$ exist and are continuous.
\end{itemize}
Roughly speaking, Assumptions \textbf{B1}-\textbf{B4} are multivariate versions of \textbf{A1}-\textbf{A4}. For simplicity, assumption \textbf{B1} assumes that the covariates are supported on the unit interval. However, it can be supported in any other compact set, with slight adaptation in the proof to account for that.    \textbf{B2} restricts the analysis to compactly supported Lipschitz kernel functions. \textbf{B3} gives a slightly stronger  unconditional moment bound in comparison with \textbf{A3}.
In what follows, we establish the asymptotic normality and derive the uniform convergence rate in $\bs x$  for the SBS. Before we proceed, we need some notation. Let $\sigma^2_j(x)=\var\big(Z-g(\bs{X})|X_j=x\big)$ and let $\mathscr{A}_{[0,1]^m}$ denote the space of real functions defined in $[0,1]^m$ with components $\beta_j:[0,1]\rightarrow \R$, satisfying
\begin{equation*}
\int_0^1 \beta_j(x_j)f_{X_j}(x_j)dx_j=0, \quad \mbox{ for }\ j\in\{1,\cdots,m\},
\end{equation*}
$\beta_0$ being constant, and denote by 
\begin{align}\label{eqaux1}
(\hat\beta_0,\ldots,\hat\beta_m)
&=\argmin_{(\beta_0,\ldots,\beta_m)\in\R\times\mathscr{A}_{[0,1]^{m}}}
\Bigg\{
\int_{[0,1]^m}
    \big(\beta(\bs{x}) - \beta_0 - \beta_1(x_1) - \cdots - \beta_m(x_m)\big)^2 \notag\\
&\qquad\times f_{\bs{X}}(\bs{x})\, d\bs{x}
\Bigg\},
\end{align}
for a given function $\beta:[0,1]^m\rightarrow \R$.

 \begin{thm}\label{teo2.2}
Suppose that \textbf{B1}-\textbf{B4} hold and that $n^{1/5}\bs h\to \bs c_{ h}$, as $n\to\infty$, for some  $\bs c_{h}\in\R_+^m$ with strictly positive entries. Then, for all $j\in\{1,\cdots,m\}$,
\begin{align*}
\max_{x_j\in[0,1]} \Big\{\big|\hat g_j(x_j)-g_j(x_j)\big|\Big\}&=O_p\big(n^{-1/5}\big),\\
\max_{x_j\in[h_j,1-h_j]} \Big\{\big|\hat g_j(x_j)-g_j(x_j)\big|\Big\}&=O_p\big(n^{-2/5}\big),
\end{align*}
and thus,
\begin{align}\label{equasb26}
\max_{\bs x\in[0,1]^m} \Big\{\big\lvert\hat g(\bs x)-g(\bs x)\big\rvert\Big\}&=O_p\big(n^{-1/5}\big),\\\label{equasb26c}
\max_{\bs x\in[\bs h,1-\bs h]} \Big\{\big\lvert\hat g(\bs x)-g(\bs x)\big\rvert\Big\}&=O_p\big(n^{-2/5}\big),
\end{align}
 where $[\bs h,1-\bs h]=[h_1,1-h_1]\times \cdots\times [h_m,1-h_m]$. Moreover,  for any $\bs{x}\in[\bs h,1-\bs h]$,
\begin{equation}\label{equasb27}
n^{2/5}\left[
\begin{array}{c}
       \hat g_1(x_1)-g_1(x_1)\\
     \vdots\\
       \hat g_m(x_m)-g_m(x_m)
\end{array}
\right]
\overset{d}{\longrightarrow} N\Big(\diag(\bs c_h)^2\bs{B}(\bs{x}),\bs V(\bs{x})\Big),
\end{equation}
where $\bs{B}(\bs{x})=(\hat\beta_1(x_1),\cdots,\hat\beta_m(x_m))^\intercal$, $\bs V(\bs{x})=\diag\{v_1(x_1),\cdots,v_m(x_m)\}$,
\begin{equation*}
v_j(x_j)=\frac{\sigma^2_j(x_j)}{c_{h_j} f_{X_j}(x_j)}\int_{-1}^1 K(u)^2du,
\end{equation*}
and $\hat\beta_j$ defined by  \eqref{eqaux1}, $j\in\{1,\cdots,m\}$, with
\begin{equation*}
    \beta(\bs{x})=\sum_{j=1}^m\bigg[
    \frac{g_j'(x_j)}{f_{\bs{X}}(\bs{x})}\frac{\partial f_{\bs{X}}(\bs{x})}{\partial x_j}+\frac{g_j''(x_j)}{2}
    \bigg]\int_{-1}^1 u^2K(u)du.
\end{equation*}
Consequently,
\begin{equation*}
    n^{2/5}\big[\hat g(\bs{x})-g(\bs{x})\big]\overset{d}{\longrightarrow} N\bigg(\sum_{j=1}^mc_{h_j}^2\beta_j(x_j),\sum_{j=1}^m v_j(x_j)\bigg).
\end{equation*}
\end{thm}
The function $\beta(\bs{x})$ corresponds to the asymptotic bias of the full-dimensional Nadaraya-Watson smoother which is well known to depend on the design density $f_{\bs{X}}$. Explicit solutions for functions $\hat\beta_1,\cdots,\hat\beta_m$ are non-trivial. Note that Theorem \ref{teo2.2} shows the pointwise convergence of $\tilde g_j(x_j)-g_j(x_j)=O_p(n^{-2/5})$ on interior points $x_j\in(0,1)$ which contrasts with the  $O_p(n^{-1/5})$  rate typically observed at the boundary points, due to the well-known bias issues of Nadaraya-Watson estimators. The smooth backfitting can alternatively be  constructed based on local linear estimates which avoids such boundary effects \citep[see Theorem 6.2 of][]{mammen_band}. Also, its asymptotic bias does not depend on the design density and it has the oracle property \citep[see Theorem 4' of][]{mammen}. However,  the local linear smooth backfitting is considerably more complicated to implement.

\begin{thm}\label{teo3}
Suppose that assumptions  \textbf{B1}-\textbf{B4} hold and  that $n^{1/5}\bs h\to \bs c_{ h}$, as $n\to\infty$, for some  $\bs c_{h}\in\R_+^m$ with strictly positive entries. Then,
\begin{equation*}
\hat p -p=O_p\bigg(\frac{1}{\sqrt{n}}\bigg),\qquad\mbox{ and }\qquad \sqrt{n}(\hat p-p)\overset{d}{\longrightarrow} N(0,3p^2/2).
\end{equation*}
Moreover,
\begin{equation*}
\max_{\bs x\in[0,1]^m}\big\{\big|\hat f(\bs x)-f(\bs x)\big|\big\}=O_p\big(n^{-1/5}\big),
\end{equation*}
and for any $\bs{x}\in[\bs h,1-\bs h]$,
\begin{equation*}
    n^{2/5}\big[\hat f(\bs{x})-f(\bs{x})\big]\overset{d}{\longrightarrow} f(\bs x) N\bigg(\sum_{j=1}^mc_{h_j}^2\beta_j(x_j),\sum_{j=1}^m v_j(x_j)\bigg).
\end{equation*}
\end{thm}
The result in Theorem \ref{teo3} shows that $\hat p$ is an oracle $\sqrt{n}$-consistent estimator of $p$ and $\hat f$ is dominated by the nonparametric part, the same conclusion as the case of one-dimensional input.
\subsection{ Estimating the efficiency with order statistics}
Full nonparametric methods, such as those of \cite{martins} and \cite{martins2013}, rely on the assumption that there exists at least one observed production unit that is efficient, or nearly efficient, thereby motivating the estimation of the efficiency component using the sample maximum. Although $R_i=1$ has probability zero when $R_i$ follows the Matsuoka distribution, the efficiency parameter $p$ can still be estimated through certain order statistics, with asymptotic motivation.
\begin{lemma}\label{lem1}
    If $R_i\sim M(p)$, then $\displaystyle{\min_{1\leq i\leq n}}\{-\ln (R_i)\}=O_p(n^{-2/3})$.
\end{lemma}
Recall from Section \ref{pf} that $\epsilon_i=-\ln(R_i)-3/(2p)$. From Lemma \ref{lem1}, it holds that  $\displaystyle{\min_{1\leq i\leq n}\{-\ln(R_i)\}}=o_p(1)$, which leads to the estimator
\begin{equation}\label{eqpm}
    \hat p_m=-\frac{3}{2\displaystyle{\min_{1\leq i\leq n}\{\hat\epsilon_i\}}}.
\end{equation}
The following result states the convergence rates  of the estimator $\hat p_m$.
\begin{thm}\label{teo36}
    If \textbf{A1}-\textbf{A5} hold, then
    \begin{equation*}
        \hat p_m-p=O_p\bigg(h^2+\sqrt{\frac{\ln(n)}{nh}}\bigg).
    \end{equation*}
    Moreover, if \textbf{B1}-\textbf{B4} hold and $n^{1/5}\bs h\to \bs c_h$, as $n\to\infty$, for some  $\bs c_{h}\in\R_+^m$ with strictly positive entries, then
   \begin{equation*}
        \hat p_m-p=O_p\big(n^{-1/5}\big).
    \end{equation*}
    In both settings, $P(\hat p_m>0)\to 1$ as $n\to\infty$.
\end{thm}
Theorem \ref{teo36} shows that the estimator \eqref{eqpm} inherits the convergence rates of the first-step nonparametric estimators which are clearly slower than $n^{-1/2}$, the parametric rate attained by the method of moments estimator $\hat p$. Moreover, as $\hat p$ depends on the entire sample, it is naturally more robust to outliers than $\hat p_m$.

To increase the robustness of the estimation,  \cite{chen} proposed a method based on the $\alpha_n-$quantiles of the sample (instead of the sample maximum).  Following their ideas, we take the population $\alpha_n-$quantile of $\epsilon=-\ln(R)-3/(2p)$ yielding
\begin{equation}\label{eqquant}
    Q_{n,\epsilon}=Q_{n,-\ln(R)}-\frac{3}{2p}=\frac{\gamma_n}{p}-\frac{3}{2p},\quad \forall n\in\N,
\end{equation}
where $Q_{n,\xi}\coloneqq \inf\big\{x\in\R:P(\xi\leq x)\geq \alpha_n\big\}$ for any random variable $\xi$, and
\begin{equation*}
    \gamma_n\coloneqq \gamma^{-1}\bigg(\frac{3}{2},\frac{\alpha_n\sqrt{\pi}}{2}\bigg)
\end{equation*}
is the inverse lower incomplete gamma function defined by $y=\gamma^{-1}(3/2,z)$ and $\gamma(3/2,y)=z$ for all $z\in (0,\Gamma(3/2))$. Equation \eqref{eqquant} motivates the estimation of $p$ using
\begin{equation}
    \hat p_\alpha=\frac{\gamma_n-3/2}{\hat Q_{n,\hat \epsilon}},
\end{equation}
where $\hat Q_{n,\xi}\coloneqq (1-t)\xi_{(k)}+t\xi_{(k+1)}$ with $k=\lfloor r\rfloor$, $r=(n-1)\alpha_n+1$, $t=r-k$, is a sample $\alpha_n-$quantile for any sample $\{\xi_i\}_{i=1}^n$. In  Corollary 3.1 of \cite{chen},
it is assumed that $\alpha_n\to 1$  with $n(1-\alpha_n)$ diverging to infinity at an appropriate rate as $n\to\infty$. In our setting, where the minimum rather than the maximum is replaced by the $\alpha_n$-quantile, we assume instead that $\alpha_n\to 0$ and $n\alpha_n\to\infty$. Note that this excludes the case $\alpha_n=0$. On the other hand, if $\alpha_n=0$, then $\hat Q_{n,\hat \epsilon}=\displaystyle{\min_{1\leq i\leq n}}\{\hat \epsilon_i\}$ and $\gamma_n=0$ \cite[Equation 8.350.5 of][]{grad}, so that $\hat p_\alpha$ coincides with $\hat p_m$.
\begin{thm}\label{teo37}
    Suppose that \textbf{A1}-\textbf{A5} hold. If $\alpha_n\to0$, $n\alpha_n\to\infty$ and $n^vh\to c_h$ for some $c_h>0$ with $v\in(0,1)$, then
    \begin{equation*}
        \hat p_\alpha-p=O_p\bigg(h^2+\sqrt{\frac{\ln(n)}{nh}}\bigg).
    \end{equation*}
    Moreover, if \textbf{B1}-\textbf{B4} hold and $n^{1/5}\bs h\to \bs c_h$, for some  $\bs c_{h}\in\R_+^m$ with strictly positive entries, then
   \begin{equation*}
        \hat p_\alpha-p=O_p\big(n^{-1/5}\big).
    \end{equation*}
     In both settings, $P(\hat p_\alpha>0)\to 1$ as $n\to\infty$.
\end{thm}
Theorem \ref{teo37} shows that the $\alpha$-quantile estimator $\hat p_\alpha$ behaves similarly as $\hat p_m$: its convergence is dominated by the nonparametric estimation.
Due to these results, we will focus exclusively on the method of moments estimator $\hat p$.
\section{Simulations}\label{sim}
In this section, we investigate the finite sample behavior of the estimators presented in Section \ref{seca} via Monte Carlo experiments.
We simulate model \eqref{equa1} through the following data-generating process (DGP) considering both univariate and bivariate input settings:
\begin{itemize}
    \item[(i)] Random samples are generated from $Y_i=f(X_i)R_i$, for all $i\in\{1,\cdots,n\}$, with $ f(X_i)=-X_i^2+4X_i$, and $X_i\sim U(1,2)$ being independent of $R_i\sim M(p)$;
    \item[(ii)]Random samples are generated from $Y_i=f(X_i)R_i$, for all $i\in\{1,\cdots,n\}$, with $ f(X_i)=\exp(-f_1(X_{i,1})-f_2(X_{i,2}))$, $f_1(x)=-1.5x^2+3x-1$, $f_2(x)=-(\ln(x)+1)/2+\ln(2)$, and $X_{i,1}\sim U(1,2)$, $X_{i,2}\sim U(1,2)$ and $R_i\sim M(p)$ being mutually independent.
\end{itemize}
The simulation is replicated $N=1{,}000$ times for each combination of sample size $n\in\{50,100,250\}$ and parameter $p\in\{1,2,8\}$.

The performance of the estimators $\hat f$ and $\hat g$ is assessed by the mean average squared error (MASE).
Let $\hat m$ be an estimator of $m$, and suppose $N$ replications are available. For a realization $(x_{1,r},\cdots,x_{n,r})$ from $\bs X$,  $\ase_r(\hat m)\coloneqq  n^{-1}\sum_{i=1}^n (\hat m(x_{i,r})-m(x_{i,r}))^2$ is computed for each $r\in\{1,\cdots,N\}$. Accordingly, the MASE is defined as the average ASE across all replications,  $L(\hat m)\coloneqq N^{-1}\sum_{r=1}^N ASE_r(\hat m)$. The simulation was performed using the software R version 4.2.3 \citep{R}. The code was implemented by the authors and is available on \href{https://github.com/marciovalk/Frontier-Estimation---Matsuoka-distribution}{github.com/marciovalk/Frontier-Estimation---Matsuoka-distribution}. The incomplete and inverse incomplete gamma functions were calculated using the \texttt{zipfR} library \citep{zipfr}.

The nonparametric estimator employed in the first step of our estimation procedure requires the choice of a kernel function and a bandwidth (also called smoothing parameter). Here, simulations were conducted using the Epanechnikov kernel. The bandwidth selection is usually done by a \textit{cross-validation} algorithm or a \textit{plug-in} method \citep[see][]{wand_jones,fan_gijbels}. In this study, we focus on the \textit{leave-one-out}  cross-validation method that minimizes the following loss function on a set $H\subseteq\R^m$ conditionally on $\bs{X}=(\bs{X}_1,\cdots,\bs{X}_n)$:
\begin{equation*}
    CV(\bs{h}|\bs{X})=\frac{1}{n}\sum_{i=1}^n (Z_i-\hat g_{-i}(\bs{X}_i))^2,
\end{equation*}
where $\hat g_{-i}$ is the estimate of $g$ calculated from the subsample $\big\{(Y_j,\bs{X}_j), j\in\{1,\cdots, n\}\backslash\{i\}\big\}$ for all $i\in\{1,\cdots,n\}$ with $\bs{X}_j\in\R^m$. We define the leave-one-out bandwidth as
\begin{equation*}
    \bs{h}_{cv}=\argmin_{\bs{h}\in H}\{CV(\bs{h}|\bs{X})\}.
\end{equation*}
We compare our production frontier estimates with those estimated by the full nonparametric method of \cite{chen}, which will be denoted as $\hat f_{\text{ctz}}$. For a better comparison, the only component in $\hat f_{\text{ctz}}$ that differ from our procedure is the estimation of the mean efficiency which is calculated nonparametrically by the sample maximum.
\subsubsection*{DGP (i): univariate inputs}
Simulation results for univariate inputs are summarized in Table \ref{tab:sim}, which reports MASE values for estimators $\hat f$ and $\hat g$, and expected values, variances, and the 0.05 and 0.95 quantiles for $\hat p$. Observe that  the MASE and $\var(\hat p)$ values decrease toward zero as $n$ increases, while the expected value of $\hat p$ approaches $p$. This provides evidence that the estimates uniformly converge to their respective true values, in line with the theory develop in Section \ref{at}.

\begin{table}[ht]
\centering
\caption{Monte Carlo simulation results for univariate inputs.}\label{tab:sim}
\vskip.2cm
\begin{tabular}{c|c|ccc|cccc}
  \hline
 $p$&$n$&$L(\hat f)$  & $L(\hat f_{\text{ctz}})$  &$L(\hat g)$  & $\E(\hat p)$ & $\var(\hat p)$& $Q_{0.05}(\hat p)$&$Q_{0.95}(\hat p)$ \\
  \hline
\multirow{3}{*}{1}& $50$&1.30 & 2.11 & 0.09 & 1.10 & 0.04 & 0.83 & 1.44 \\
 & $100$&    0.62 & 1.23 & 0.04 & 1.05 & 0.02 & 0.85 & 1.25 \\
 & $250$ & 0.27 & 0.52 & 0.02 & 1.02 & 0.01 & 0.89 & 1.15 \\
\hline
\multirow{3}{*}{2}&  $50$&0.28 & 0.34 & 0.02 & 2.16 & 0.13 & 1.63 & 2.81 \\
 & $100$ &0.16 & 0.18 & 0.01 & 2.09 & 0.07 & 1.69 & 2.53 \\
 & $250$ &0.07 & 0.08 & 0.00 & 2.04 & 0.02 & 1.79 & 2.29 \\
  \hline
 \multirow{3}{*}{8}& $50$&0.02 & 0.02 & 0.00 & 8.67 & 2.43 & 6.35 & 11.44 \\
 & $100$ &0.01 & 0.01 & 0.00 & 8.36 & 1.13 & 6.76 & 10.21 \\
 & $250$ & 0.01 & 0.01 & 0.00 & 8.11 & 0.38 & 7.10 & 9.12 \\
   \hline
\end{tabular}
\end{table}

\begin{figure}[ht]
\centering
\includegraphics[width=0.8\textwidth]{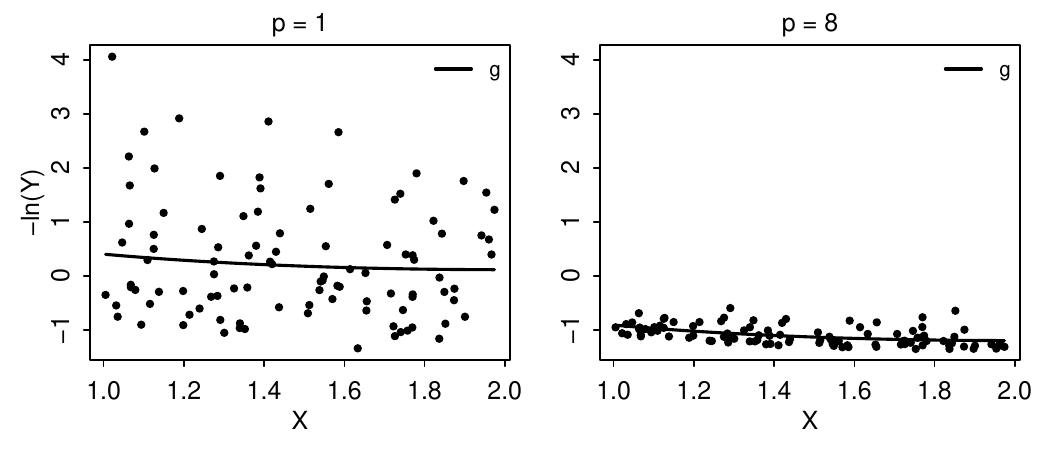}
\caption{ Simulated data for $p=1$ and $p=8$.}\label{fig:sim_p}
\end{figure}

Table \ref{tab:sim} suggests that the MASE values for $\hat g$ and $\hat f$ tend to be larger when $p$ is small.  Figure \ref{fig:sim_p} provides a visual presentation of $n=100$ simulated data points from DGP (i) for $p\in\{1,8\}$. The plots indicate that the variance of the error term is larger when $p=1$. For instance, $ \sigma^2_{\epsilon}\simeq 1.5$ for $p=1$, whereas $ \sigma^2_{\epsilon}\simeq 0.02$ for $p=8$.
In view of Theorem \ref{teo1},  this is expected behavior, since the variance of $\hat g(x)$, given by $3\int_\R K(u)^2du/(2p^2f_X(x)nh)$, depends inversely on $p$, while its bias is $p$-independent. Moreover, Theorem \ref{teo2.2} shows that the estimation of $f$ is dominated by the first-step nonparametric part, meaning that $\hat f$ tends to perform well provided that $\hat g$  performs well. This links the accuracy of $\hat f$ and $\hat g$ positively, as evidenced in Table \ref{tab:sim}.

\begin{figure}[ht]
\centering
\includegraphics[width=0.8\textwidth]{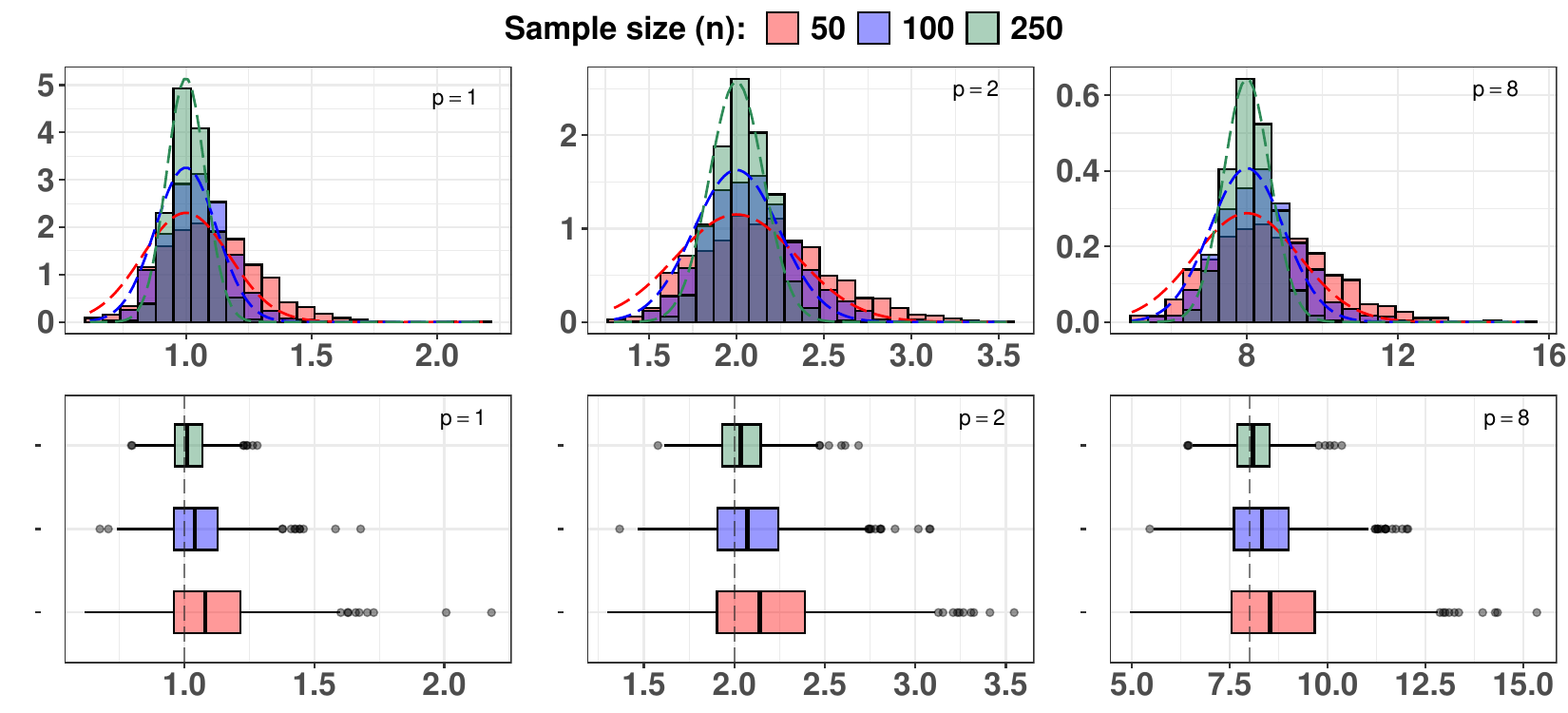}
\caption{Boxplots of the estimated values $\hat p$ for the univariate case.}\label{fig:sim_uni}
\end{figure}

\begin{figure}[ht]
\centering
\includegraphics[width=0.8\textwidth]{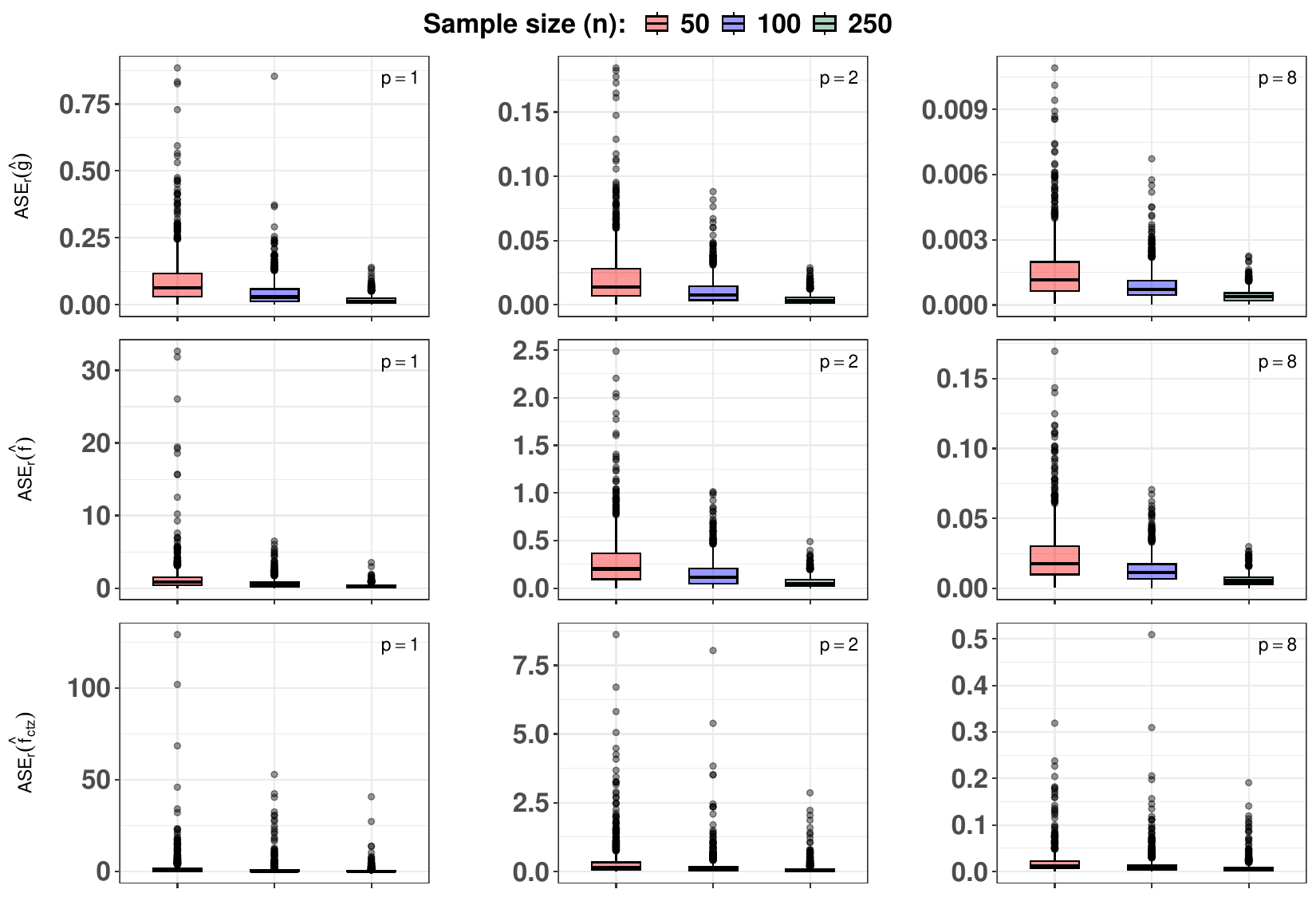}
\caption{Boxplots of the values $\ase_r(\hat g)$ and $\ase_r(\hat f)$ for the univariate case.}\label{fig:sim_uni_ase}
\end{figure}

Under mild conditions, it can be shown  \citep[Lemma 4.2 in][]{xia} that the averaged square error $\ase(\hat g)\coloneqq n^{-1}\sum_{i=1}^n (\hat g(X_{i})-g(X_{i}))^2$ is asymptotically equivalent to the mean integrated squared error (MISE) defined by $\int \E[(g(x)-\hat g(x))^2]f_X(x)dx=\int [\bias(\hat g(x))^2+\var(\hat g(x))]f_X(x)dx$ which explains the positive relationship between $L(\hat g)$ and $\var(\hat g)$. If smaller values of $p$ tend to increase $\ase(\hat g)$ via $\var(\hat g)$, then they also tend to increase $\E\big(\ase(\hat g)\big)$, which is approximated by $L(\hat g)$. 
Table \ref{tab:sim}  indicates that the variance of $\hat p$ depends positively on $p$.
This is consistent with Theorem \ref{teo2}, which shows that the asymptotic variance of $\hat p$ is $3p^2/(2n)$, and therefore positively related to $p$. As for the bias, Table \ref{tab:sim}  suggests that $\hat p$ inherits the fact that the unfeasible estimator $\tilde p$ overestimates $p$ which can be seen through repeated applications of Jensen's inequality as follows
\begin{align*}
    \E(\tilde p)>\sqrt{\frac{3n}{2}}\frac{1}{\E\Big (\sqrt{\sum_{i=1}^n \epsilon_i^2}\Big )}\geq \sqrt{\frac{3n}{2}} \frac{1}{\sqrt{\sum_{i=1}^n E(\epsilon_i^2)}}=p.
\end{align*}
 Figure \ref{fig:sim_uni} presents boxplots and histograms of the simulation results for different values of $p\in\{1,2,8\}$ and $n\in\{100,150,250\}$. The plots show finite sample evidence of the consistency and asymptotic normality of $\hat p$ obtained in Theorem \ref{teo2}. The dashed lines in the histograms illustrate the densities of the corresponding asymptotic normal distributions $N(p,\tfrac{3p^2}{2n})$. In Figure \ref{fig:sim_uni_ase}, we observe that $\ase_r(\hat f)$ and $\ase_r(\hat g)$ decrease to values close to zero as the sample size increases. Aside from the scale, we observe that the behavior of $\ase_r(\hat f)$ and $\ase_r(\hat g)$ are quite similar.

Regarding the estimator $\hat f_{\text{ctz}}$ of \cite{chen}, Table \ref{tab:sim} and Figure \ref{fig:sim_uni_ase} indicate that it behaves as a consistent estimator of $f$, but with a poorer performance compared to our estimator $\hat f$. Nevertheless, its performance can still be considered satisfactory, given that it is a more general procedure not specifically tailored to efficiency estimation under the Matsuoka distribution.

\subsubsection*{DGP (ii): bivariate inputs}
\begin{table}[ht]
\centering
\caption{Monte Carlo simulation results for the case of bivariate inputs - CBS.}\label{tab:sim2}
\vskip.2cm
\begin{tabular}{c|c|ccc|cccc}
  \hline
$p$ & $n$&$L(\hat f)$ &$L(\hat f_{\text{ctz}})$ & $L(\hat g)$ & $\E(\hat p)$ & $\var(\hat p)$& $Q_{0.05}(\hat p)$&$Q_{0.95}(\hat p)$ \\
  \hline
\multirow{3}{*}{1}& $50$&0.27 & 0.68 & 0.16 & 1.15 & 0.04 & 0.84 & 1.50 \\
 &$100$&  0.16 & 0.46 & 0.08 & 1.07 & 0.02 & 0.86 & 1.30 \\
 &$250$ & 0.07 & 0.21 & 0.04 & 1.03 & 0.01 & 0.90 & 1.16 \\
  \hline
\multirow{3}{*}{2}&   $50$& 0.08 & 0.11 & 0.04 & 2.28 & 0.17 & 1.68 & 3.03 \\
 &$100$ &0.04 & 0.08 & 0.02 & 2.14 & 0.07 & 1.70 & 2.58 \\
 &$250$ &0.02 & 0.05 & 0.01 & 2.06 & 0.03 & 1.80 & 2.34 \\
  \hline
\multirow{3}{*}{8}&   $50$& 0.01 & 0.01 & 0.00 & 9.16 & 2.98 & 6.70 & 12.26 \\
  &$100$ &  0.00 & 0.01 & 0.00 & 8.60 & 1.27 & 6.82 & 10.55 \\
  &$250$ &   0.00 & 0.00 & 0.00 & 8.23 & 0.39 & 7.26 & 9.27 \\
   \hline
\end{tabular}
\end{table}
Simulation results for the estimation based on CBS and SBS estimators are summarized in Tables \ref{tab:sim2} and \ref{tab:sim3}, respectively. Histograms and boxplots for $\hat p, \ase_r(\hat g), \ase_r(\hat f)$ and $\ase_r(\hat f_{\text{ctz}})$ can be found in the supplementary material (Figures 5-8). Both procedures performed quite similarly and the general conclusions are the same as in the univariate case: (a) they indicate that the estimates get closer to their respective true values as the sample size increases; (b) the MASE values for $\hat g$ and $\hat f$ are larger for smaller values of $p$; (c) $\hat p$ tends to overestimate $p$, and its bias and variance increase with $p$; and the full nonparametric estimate $\hat f_{\text{ctz}}$ performs worse than $\hat f$.
\FloatBarrier
\begin{table}[hb]
\centering
\caption{Monte Carlo simulation results for the case of bivariate inputs - SBS.}\label{tab:sim3}
\vskip.2cm
\begin{tabular}{c|c|ccc|cccc}
  \hline
$p$ & $n$&$L(\hat f)$&$L(\hat f_{\text{ctz}})$   & $L(\hat g)$ &  $E(\hat p)$ & $\var(\hat p)$& $Q_{0.05}(\hat p)$&$Q_{0.95}(\hat p)$ \\
  \hline
\multirow{3}{*}{1}& $50$&0.19 & 0.35 & 0.13 & 1.10 & 0.04 & 0.81 & 1.45 \\
 &$100$&  0.13 & 0.26 & 0.08 & 1.05 & 0.02 & 0.85 & 1.29 \\
 &$250$ &0.07 & 0.14 & 0.04 & 1.02 & 0.01 & 0.90 & 1.16 \\
  \hline
\multirow{3}{*}{2}&   $50$& 0.08 & 0.10 & 0.04 & 2.25 & 0.19 & 1.63 & 3.02 \\
 &$100$ &0.05 & 0.08 & 0.02 & 2.13 & 0.07 & 1.73 & 2.60 \\
 &$250$ &  0.03 & 0.05 & 0.01 & 2.04 & 0.03 & 1.78 & 2.31 \\
  \hline
\multirow{3}{*}{8}&   $50$&0.01 & 0.01 & 0.01 & 9.15 & 3.08 & 6.51 & 12.29 \\
  &$100$ & 0.01 & 0.01 & 0.00 & 8.59 & 1.31 & 6.86 & 10.54 \\
  &$250$ & 0.00 & 0.01 & 0.00 & 8.29 & 0.47 & 7.24 & 9.56 \\
   \hline
\end{tabular}
\end{table}

\section{Application to agricultural data}\label{ea}

In Brazil, temporary crops represent a substantial share of agricultural output. According to the 2017 Agricultural Census \cite{censoagro}, within total agricultural production, 77\% was generated by temporary crops. Within this segment, the South and Center-West regions stood out as the two largest producers in the country, contributing 29.07\% and 36.12\% of total production value, respectively. While the South is characterized by smaller farms, greater diversification, and stronger cooperative networks, the Center-West is distinguished by larger production units, a stronger orientation toward export commodities, and the predominance of monoculture systems.

These structural contrasts are also reflected in broader statistics on the agricultural and livestock sector in 2017: the total area of establishments amounted to 42.9 million hectares in the South compared to 112 million hectares in the Center-West; moreover, 54.15\% of all cooperatives in Brazil were located in the South, whereas only 7.96\% were in the Center-West. Family farming was likewise more prevalent in the South, which accounted for 17.08\% of all family farming establishments in Brazil, compared to 5.73\% in the Center-West.

To investigate how these structural differences translate into production technology, we apply our two-step estimation procedure to data on Brazilian temporary crops from the 2017 Agricultural Census \citep{censoagro}. The dataset comprises $n_s = 84$ observations for the South region and $n_c = 50$ observations for the Center-West.\footnote{The initial dataset comprises 94 microregions for the South and 51 for the Center-West. Based on \cite{chen}, we inspected the residuals $\hat \epsilon$, and, using the inter-quartile procedure proposed by \cite{fang}, identified a few outliers, which were subsequently discarded from the analysis.} The \textit{gross value of production} ($Y$) is taken as the response variable, while the \textit{amount of land area utilized for temporary crops} ($X_1$) and the \textit{expenditure on fertilizers and pesticides} ($X_2$) serve as covariates. In this setting, the decision-making unit is the microregion, and all variables are normalized by the total number of production units in the corresponding microregion. For both South and Center-West regions, we estimate the respective production frontiers, $f_s$ and $f_c$, and the respective efficiency parameters $p_s$ and $p_c$ using model \eqref{equa1}. The estimation is performed according to the procedure in Section \ref{seca} with the nonparametric SBS estimator. Bandwidths are selected via leave-one-out cross-validation described in Section \ref{sim} and smoothing was done using the Epanechnikov kernel.
\begin{figure}[ht]
\centering
\includegraphics[width=0.8\textwidth]{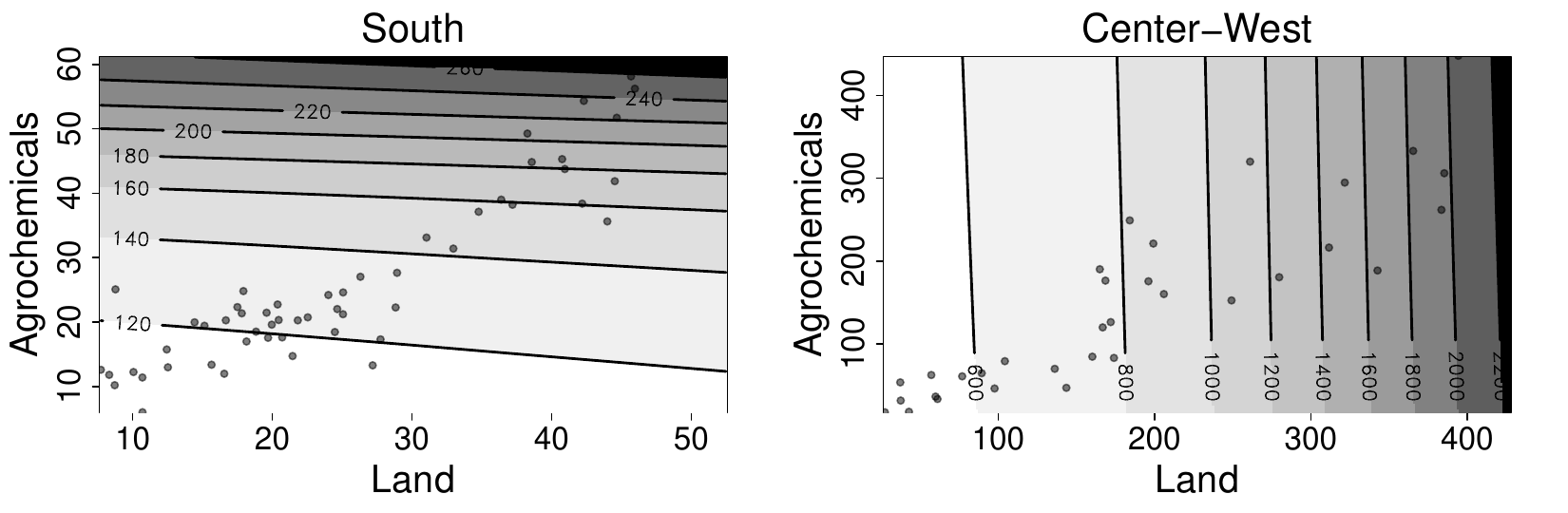}
\caption{Contour lines of the estimated production functions $\hat f_s$ and $f_c$.}\label{fig:aplic}
\end{figure}
Figure \ref{fig:aplic} shows the isoquants obtained from the estimated production frontier $\hat f(x_1,x_2)$ over the relevant range of observed input values.  The results indicate contrasts between the two regions, but also a common feature: in both cases, the estimated isoquants are close to the constant elasticity of substitution (CES) form, consistent with production technologies that display nearly constant substitution rates between inputs.

In the South, the estimated isoquants are nearly horizontal, indicating that agrochemicals have higher relative marginal productivity. In other words, agrochemicals are the critical input: sustaining production requires their use, since reductions in agrochemicals cannot be easily compensated by additional land. This configuration is characteristic of an intensive production system, in which scarce land is cultivated with greater reliance on agrochemicals intensification to sustain yields. It is consistent with the structural features of Southern agriculture, marked by smaller farms, limited land availability, and diversified practices that prioritize productivity gains per unit of area.

In the Center-West, the isoquants are nearly vertical. This shape indicates a high marginal rate of technical substitution of land for agrochemicals: sizable reductions in agrochemical use can be offset by only small increases in cultivated area, so production depends primarily on land availability. This pattern is consistent with an extensive production regime built on abundant land and large-scale monoculture, with agrochemicals playing a supportive role.

The estimated values of $p_s$ and $p_c$ are $\hat p_s= 4.71$ and $\hat p_c= 2.98$, which suggest that the efficiency distribution of both regions are left-skewed. Note that the efficiency $R$ follows the Matsuoka distribution with parameter $p$ if, and only if,  $\epsilon=-\ln(R)-3/(2p)$ has a shifted Gamma distribution with cumulative distribution function
\begin{equation}\label{distgamma}
	F_\epsilon(x)=\frac{1}{\sqrt\pi}\gamma\big(1.5,xp+1.5\big) I(x>3/(2p)),
\end{equation}
where $\gamma(s,x)$ denotes the lower incomplete gamma function. This distribution admits the quantile function
\begin{equation}\label{distgammainv}
	F_\epsilon^{-1}(q)=\frac{1}{p}\bigg(\gamma^{-1}\big(1.5,q\sqrt\pi\big)-\frac{3}{2}\bigg), \quad q\in(0,1),
\end{equation}
where $\gamma^{-1}(s,y)$ denotes the inverse of $\gamma(s,x)$.

To assess the plausibility of the Matsuoka distributional assumption, we performed Chi-squared goodness-of-fit tests under the null that the residuals $\hat \epsilon$ follow the distribution in \eqref{distgamma} with $p\in\{4.71, 2.98\}$. This test is asymptotically valid as $\hat \epsilon$ converges in probability to $\epsilon$ with a rate depending on the first-step nonparametric estimator ($O_p(n^{-1/5})$ for SBS). Since  \eqref{distgammainv} determines the quartiles of \eqref{distgamma}, four bins were constructed accordingly. The resulting p-values were $0.40$ for the South and $0.82$ for the Center-West (under the null, we used 2 degrees of freedom for the Chi-square distribution to account for the estimation of $p$). Thus, for both regions individually, we find no strong evidence against the null that the efficiency distribution follows the Matsuoka law with the corresponding estimated parameter.

\begin{figure}[ht]
	\centering
	\includegraphics[width=0.8\textwidth]{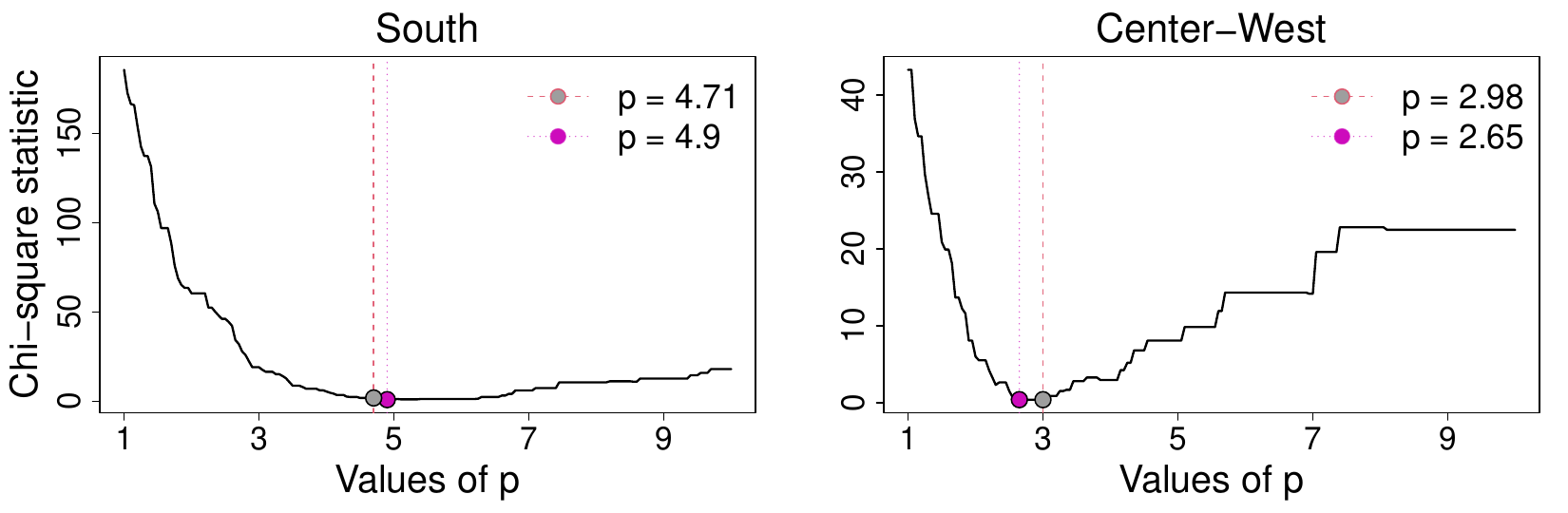}
	\caption{Observed Chi-square statistics for different values for the parameter $p$.}\label{fig:chi_stats}
\end{figure}

To complement the residual diagnostics, we evaluate the observed chi-squared statistic for each region on a  grid of 181 equally spaced points of $p$ of the form $\{p>0:p=1+0.05k\}_{k=0}^{180}$. Figure \ref{fig:chi_stats} shows that the estimated values of $p$ ($\hat p_s=4.71$ and $\hat p_c=2.98$) lie quite close to the minimum points of the observed chi-squared statistics, which occur at $p=4.9$ and $p= 2.65$ for South and Center-West, respectively. This supports the adequacy of our estimators $\hat p_s$ and $\hat p_c$, indicating that they provide a good fit for the dataset.
Because the efficiency distribution depends solely on $p$, a central question is whether $p_s=p_c$. Standard inference to test this requires, at a minimum, approximate independence between the regional samples. Since the South and Center-West are contiguous regions, severe spatial dependence could undermine the validity of traditional inference. To address this concern, we apply the spatial autocorrelation test using the Moran's I statistics \citep{moran} to observations located along the border between the two regions, considering microregions that lie on the frontier as well as those within two neighboring layers from it\footnote{The test was implemented using the \texttt{moran.mc} function from the R package \texttt{spdep}, with significance assessed through 2000 Monte Carlo replications. The shapefile containing the georeferencing information was obtained from \cite{ANA}.}. The resulting p-value  is 0.18, providing no strong evidence against the null of spatial independence. However, independence cannot be formally proven and other sources of dependence may exist. We therefore proceed under the assumption of approximate independence, while noting that inferential conclusions should be interpreted with caution.

According to Theorem \ref{teo3}, the null $H_0: p_s=p_c$ can be tested with
\begin{equation}
    Z_0=\frac{\hat p_s-\hat p_c}{\hat\sigma_{s,c}}\overset{a}{\sim} N(0,1), \text{ where } \hat\sigma_{s,c}=\sqrt{\frac{3\hat p_s^2}{2n_s}+\frac{3\hat p_c^2}{2n_c}}.
\end{equation}
For our data, the p-value for the two-sided test is 0.03, which falls bellow conventional significance levels. Hence, we  find sufficient evidence to reject the null of equal efficiency distributions.
To complement this inference, we also estimate the stress-strength probability
\begin{equation}\label{eqstress}
    P_R\coloneqq P(R_s>R_c)=\frac2\pi\big[(2s-1)\sqrt{s(1-s)}+\arcsin\big(\sqrt{s}\big)\big],
\end{equation}
where $R_s\sim M(p_s)$ and $R_c\sim M(p_c)$ are assumed independent efficiency variables for the South and Center-West regions, respectively and $s=p_s/(p_s+p_c)$. The closed-form expression in \eqref{eqstress} depends only on $s$ which summarizes the relative magnitudes of the two parameters. Substituting $\hat s=\hat p_s/(\hat p_s+\hat p_c)$, yields the estimator $\hat P_R$. According to Theorem \ref{teo3} and the Delta method, a two-sided $100\gamma$\%  confidence interval is
\begin{equation}
    CI(P_R;\gamma)=\bigg[\hat P_R-z_{\gamma}\hat\sigma_{P_R};\hat P_R+z_{\gamma}\hat\sigma_{P_R}\bigg]
\end{equation}
where $z_{\gamma}$ satisfies $P(Z>z_{\gamma})=(1-\gamma)/2$, $Z\sim N(0,1)$, and
\begin{equation}
   \hat\sigma_{P_R}=\sqrt{\frac{128}{3\pi^2}\bigg[\frac{\hat p_s\hat p_c}{(\hat p_s+\hat p_c)^2}\bigg]^3\bigg(\frac{1}{n_s}+\frac{1}{n_c}\bigg)}.
\end{equation}
For our dataset,  $\hat P_R=0.62$ with a $95$\% confidence interval of $[0.558;0.726]$. As this interval does not contain $0.5$, the null hypothesis $H_0:P_R=0.5$ is rejected at the 5\% significance level under the two-sided test (p-value of 0.09\%).\footnote{Since $P_R(s)$ is strictly increasing on $s\in(0,1)$, we have  $P_R(0.5)=0.5$ if, and only if, $p_s=p_c$. Thus, we have tested two equivalent null hypotheses, $P_R=0.5$  and $p_s=p_c$, with different test statistics that may behave differently in finite samples.}  Taken together, the results indicate that the South is relatively more efficient than the Center-West with respect to land and agrochemicals.

\section{Conclusion and discussion}\label{conc}

This paper develops a deterministic production frontier framework in which technical efficiency follows the Matsuoka distribution, a parsimonious one-parameter law on $(0,1)$ constructed to reflect empirical regularities of efficiency data. Building on this specification, we propose a two-step semiparametric estimation method: a nonparametric estimator for the regression component, followed by a feasible method of moments estimator for the efficiency parameter with plug-in reconstruction of the frontier. We establish  convergence rates and asymptotic normality for all components; the second-step parametric estimator enjoys an oracle property, remaining asymptotically unaffected by the preliminary nonparametric smoothing. Monte Carlo experiments document finite-sample behavior consistent with the theory and illustrate the practical gains from exploiting the parametric efficiency component relative to the fully nonparametric alternative of \cite{chen}.

An empirical application to Brazilian temporary crops (South vs. Center-West) with two inputs - \emph{land} and \emph{agrochemicals} - shows that the estimated isoquants in both regions are close to the constant elasticity of substitution (CES) form, yet they differ in the critical input: in the South, agrochemicals exhibit higher relative marginal productivity (near-horizontal isoquants), consistent with an intensive, land-constrained production profile; in the Center-West, land is the critical input (near-vertical isoquants), in line with an extensive, land-anchored regime and large-scale monoculture. Estimation of the efficiency parameter yields $\hat p_s=4.71$ and $\hat p_c=2.98$, with goodness-of-fit tests providing no strong evidence against the Matsuoka specification in either region. A formal comparison rejects equality of the efficiency parameters at conventional levels, and a stress-strength calculation indicates a probability above one-half that South outperforms Center-West in efficiency.

Overall, the results suggest that (i) modeling efficiency via the Matsuoka distribution offers a tractable and empirically credible route to frontier estimation; (ii) the proposed two-step procedure delivers strong asymptotic guarantees and favorable finite-sample performance; and (iii) in the Brazilian temporary-crops setting, the estimation of the efficiency parameter reveals statistically significant evidence that the South is relatively more efficient than the Center-West. This finding underscores the role of regional production structures in shaping efficiency outcomes and highlights the practical value of the proposed framework for comparative efficiency analysis.

\section*{Declarations}
\textbf{Conflict of interest}. On behalf of all authors, the corresponding author states that there is no conflict of interest.

\bibliographystyle{apalike}
\bibliography{bib}
\section*{Appendix A. Proofs}
\proof{teo1} The proof follows by direct applications of Theorem 10 of \cite{hansen} for the uniform rate of convergence in \eqref{equa24} and Theorem 5.2 of \cite{fan_gijbels} for the asymptotic normality in \eqref{equa24_}, and thus is omitted.

\proof{teo2.2} Firstly, we decompose the SBS estimator into the bias part and the stochastic part (mnemonically) as follows
\begin{equation}\label{eqdec1}
    \hat g_j(x_j)=\hat g_j^B(x_j)+\hat g_j^S(x_j), \quad \mbox{for all }\ j\in\{1,\cdots,m\},
\end{equation}
where, for $J\in\{B,S\}$,
\begin{align}\label{eqdec2}
    \hat g_j^J(x_j)=\tilde g_j^J(x_j)-\sum_{\substack{k=1 \\ k\neq j}}^m\int_0^1 \hat g_k^J(x_k)\frac{\hat f_{\bs{X}}(x_j,x_k)}{\hat f_{X_j}(x_j)}dx_k-\hat g_{0}^J,
\end{align}
with
\begin{align*}
\hat g_{0}^S=\frac{1}{n}\sum_{i=1}^n\epsilon_i,\quad
\hat g_{0}^B=g_0+\frac{1}{n}\sum_{i=1}^n\sum_{j=1}^m g_j(X_{i,j}),\quad
\tilde g_{j}^S(x_j)=\frac{1}{n\hat f_{X_j}(x_j)}\sum_{i=1}^nK_{h_j}(u_j,X_{i,j})\epsilon_i
\end{align*}
and
\[\displaystyle{\tilde g_{j}^B(x_j)=\frac{1}{n\hat f_{X_j}(x_j)}\sum_{i=1}^nK_{h_j}(u_j,X_{i,j})\bigg[g_0+\sum_{j=1}^m g_j(X_{i,j})\bigg]}.\]
By applying Theorem 6.1 of \cite{mammen_band}, we can expand the stochastic and bias parts, uniformly for $\bs x\in[0,1]^m$ and $\bs h\in[c_1n^{-1/5},c_2n^{-1/5}]^m$, for all $0<c_1<c_2$, respectively as
\begin{align} \label{eqdec4}
    \hat g_j(x_j)^S=\tilde g_{j}^S(x_j)+\frac{1}{n}\sum_{i=1}^n r_{ij}(x_j)\epsilon_i+o_p(n^{-1/2}),\quad\mbox{and}\quad
    \hat g_j(x_j)^B=g_j(x_j)+O_p(n^{-1/5}),
\end{align}
where $r_{ij}, 1\leq i\leq n, 1\leq j\leq m,$ are absolutely uniformly bounded functions, say $|r_{ij}|\leq C_i<\infty$, satisfying the Lipschitz condition
\begin{equation}\label{rislc}
    \lvert r_{ij}(u_j')-r_{ij}(u_j)\rvert\leq C \lvert u_j'-u_j\rvert.
\end{equation}
The term $\tilde g_{j}^S(x_j)$  is known to be of order $O_p\big(\sqrt{\ln(n)/(nh)}\big)=O_p\big(\sqrt{n^{-3/5}\ln(n)}\big)$ uniformly on $x_j\in[0,1]$ \citep[see result (110) in][]{mammen}. By noting that $\E(|\epsilon_1|)<\E(-\ln(R_1))+3/(2p)=3/p<\infty$, we use the Central Limit Theorem and the fact that $r_{i,j}$ are uniformly bounded functions, to obtain that
\begin{align}\label{eqsb28}
    \sup_{x_j\in[0,1]}\biggl\{\bigg\lvert \frac{1}{n}\sum_{i=1}^n r_{ij}(x_j)\epsilon_i\bigg\rvert\bigg\} &\leq \frac{1}{n}\sum_{i=1}^n |\epsilon_i|\sup_{x_j\in[0,1]}\{|r_{ij}(x_j)|\}\leq \frac{1}{n}\sum_{i=1}^n |\epsilon_i|\max_{i\in\{1,\cdots,n\}}\{C_i\}\\
    &\leq C\frac{1}{n}\sum_{i=1}^n |\epsilon_i|=O_p(n^{-1/2}),\quad \mbox{for all $j\in\{1,\cdots,m\}$.}
\end{align}
Thus, $\hat g_j(x_j)^S$ is asymptotically dominated by $\displaystyle{\sup_{x_j\in[0,1]}}\{\lvert \tilde g_j^S(x_j)\rvert\}$. Combining the results presented obtained so far, we conclude that
$\displaystyle{ \max_{x_j\in[0,1]} \Big\{\big\|\hat g_j(x_j)-g_j(x_j)\big|\Big\}=O_p\big(n^{-1/5}\big)}$.
%

As for result \eqref{equasb26}, after repeated  applications of the triangle inequality, we obtain
\begin{equation*}
    \max_{\bs x\in[0,1]^m} \Big\{\big\lvert\hat g(\bs x)-g(\bs x)\big\rvert\Big\}\leq \sum_{j=0}^m \max_{x_j\in[0,1]} \Big\{\big\lvert\hat g_j(x_j)-g_j(x_j)\big\rvert\Big\}=O_p\big(n^{-1/5}\big),
\end{equation*}
uniformly on $x_j\in[h_j,1-h_j]$, and \citep[see result 6.6 in Theorem 6.1 of][]{mammen_band}
\begin{equation}\label{eqdec5}
    \hat g_j(x_j)^B=g_j(x_j)+O(n^{-2/5})+o_p(n^{-2/5}).
\end{equation}
Using the same arguments as above, we can show that $\displaystyle{\max_{\bs x\in[0,1]^m} \big\{\big\lvert\hat g(\bs x)-g(\bs x)\big\rvert\big\}=O_p(n^{-2/5})}$.
Finally, the convergence in distribution \eqref{equasb27} is proved by applying Theorem 4 of \cite{mammen}. \qed

\proof{teo2}
Define $\hat W=\sum_{i=1}^n\hat \epsilon_i^2/n$, $\tilde W=\sum_{i=1}^n \epsilon_i^2/n$ and $\theta=3/(2p^2)$. Then
\begin{equation*}
    \hat W-\theta= (\hat W-\tilde W)+(\tilde W-\theta)\coloneqq T_{1,n}+T_{2,n}.
\end{equation*}
 Firstly, we focus on the term $T_{1,n}$. For all $i\in\{1,\cdots,n\}$, note that
\begin{align*}
    \hat\epsilon^2_i&=(\hat\epsilon_i-\epsilon_i)^2+2(\hat\epsilon_i-\epsilon_i)\epsilon_i+\epsilon_i^2=(g(X_i)-\hat g(X_i))^2+2(g(X_i)-\hat g(X_i))\epsilon_i+\epsilon_i^2.
\end{align*}
Therefore, we can write
\begin{equation*}
    T_{1,n}=\frac{1}{n}\sum_{i=1}^n (\hat\epsilon_i^2-\epsilon_i^2)=\frac{1}{n}\sum_{i=1}^n(g(X_i)-\hat g(X_i))^2+\frac{2}{n}\sum_{i=1}^n \big(g(X_i)-\hat g(X_i)\big)\epsilon_i\coloneqq P_{1,n}+P_{2,n}.
\end{equation*}
From Theorem \ref{teo1}, $P_{1,n}\leq \displaystyle{\max_{x\in[a,b]}\big\{|\hat g(x)-g(x)|\big\}^2=O_p\big(h^4+\ln(n)/(nh)\big)}$. We move to bound the term $P_{2,n}$. For this, recall that the local linear estimator $\hat g(X_i)$ is linear in $\bs Z$, that is,\footnote{For clarity's sake, we emphasize that the vector of weights
$(W_1(x),\cdots,W_n(x))^\intercal$ coincides with the equivalent kernel vector
$\bs w_{1,x}$ introduced in equation \eqref{equa3},  so that $\hat g(x) = \sum_{j=1}^n W_j(x) Z_j =\bs w_{1,x}^\top \bs Z$, for all $x\in\R$.}
\begin{align*}
    \hat g(X_i)=\sum_{j=1}^n W_j(X_i)Z_j=\sum_{j=1}^n W_j(X_i)g(X_j)+\sum_{j=1}^n W_j(X_i)\epsilon_j,\quad i\in\{1,\cdots,n\},
\end{align*}
where $W_{j}(x)=(nh)^{-1}e^\intercal_1 S_{n,x}^{-1}U\Big(\frac{X_j-x}{h}\Big)K\Big(\frac{X_j-x}{h}\Big)$, with $e_1=(1,0)^\intercal$, $U(z)=(1,z)^\intercal$  and
\begin{equation*}
S_{n,x}=\frac{1}{nh}\left[
\begin{array}{ll}
	\sum_{j=1}^n K \Big(\frac{X_j-x}{h}\Big) & \sum_{j=1}^n  K\Big(\frac{X_j-x}{h}\Big)  \Big(\frac{X_j-x}{h}\Big)\\ 
	\sum_{j=1}^n  K\Big(\frac{X_j-x}{h}\Big)\Big(\frac{X_j-x}{h}\Big) &  \sum_{j=1}^n  K\Big(\frac{X_j-x}{h}\Big)\Big(\frac{X_j-x}{h}\Big)^2
\end{array}
\right].
\end{equation*}
Thus, we can decompose the deviation between $g(X_i)$ and $\hat g(X_i)$ into a bias and a stochastic part as follows:
\begin{align*}
    g(X_i)-\hat g(X_i)&=\bigg\{g(X_i)-\sum_{j=1}^n W_j(X_i)g(X_j)\bigg\}-\bigg\{\sum_{j=1}^n W_j(X_i)\epsilon_j\bigg\}\coloneqq B_{n,i}^{g}-V_{n,i}^{g}.
\end{align*}
 Then, the term $P_{2,n}$ can be expressed as
 \begin{equation*}
     P_{2,n}=\frac{2}{n}\sum_{i=1}^n \epsilon_i B_{n,i}^g -\frac{2}{n}\sum_{i=1}^n \epsilon_i V_{n,i}^g\coloneqq B_{n}^{(g,\epsilon)}-V_{n}^{(g,\epsilon)}.
 \end{equation*}
 To analyze the bias term $B_{n}^{(g,\epsilon)}$, we apply a zero-order Taylor expansion of $g(X_j)$ around $X_i$ with Lagrange remainder to obtain
 \begin{equation*}
     g(X_j)=g(X_i)+g'\big(X_i+\tau_j(X_j-X_i)\big)(X_j-X_i),\quad \tau_j\in[0,1].
 \end{equation*}
 Now, it is known that the local linear weights satisfy $\sum_{j=1}^n W_{j}(X_i)=1$, $\sum_{j=1}^n W_{j}(X_i)(X_j-X_i)=0$, $\sup_{x\in\R}\big\{\sum_{j=1}^n |W_{j}(x)|\big\}\leq C$, and $W_j(x)=0$ if $|x-X_j|> h$  \citep[see Proposition 1.12 and Lemma 1.3 in][]{tsybakov}. Since  $g$ is twice continuously differentiable, the derivative $g'$ is  Lipschitz continuous. Therefore
 \begin{align*}
     |B^g_{n,i}|&=\bigg| \sum_{j=1}^n W_j(X_i)g(X_j) -g(X_i)\bigg|=\bigg| \sum_{j=1}^n W_j(X_i)\Big[g(X_j) -g(X_i)\Big]\bigg|\\
     &=\bigg| \sum_{j=1}^n W_j(X_i)\Big[g'\big(X_i+\tau_j(X_j-X_i)\big)(X_j-X_i)\Big]-\sum_{j=1}^n W_j(X_i)g'(X_i)(X_j-X_i)\bigg|\\
     &\leq \sum_{j=1}^n |W_j(X_i)||X_j-X_i|\big |g'\big(X_i+\tau_j(X_j-X_i)\big)-g'(X_i)\big|\\
     &\leq C \sum_{j=1}^n |W_j(X_i)||X_j-X_i|\big |\tau_j(X_j-X_i)\big|I\big(|X_j-X_i|\leq h\big)\\
     &\leq Ch^2 \sup_{x\in\R}\bigg\{\sum_{j=1}^n \big|W_j(x)\big|\bigg\}\leq Ch^2.
 \end{align*}
Thus,  $B^g_{n,i}=O(h^2)$, uniformly in $i\in\{1,\cdots,n\}$. Now, observe that $\{\epsilon_i\}_{i=1}^n$ is independent of  $\{X_i\}_{i=1}^n$, whereas $B_{n_i}^{g}$ is $\bs X$-measurable. Hence, conditionally to $\bs X=(X_1,\cdots,X_n)^\intercal$, $\{\epsilon_i B^g_{n,i}\}_{i=1}^n$ is a sequence of independent random variables, although not identically distributed. Therefore,  $\E\big( B_{n}^{(g,\epsilon)}|\bs X\big)=2/n\sum_{i=1}^n \E\big(\epsilon_i B_{n,i}^g|\bs X)=2/n\sum_{i=1}^n \E(\epsilon_i) B_{n,i}^g=0$ and
\begin{align*}
    \var\Big(B_{n}^{(g,\epsilon)}|\bs X\Big)&=\frac{4}{n^2}\sum_{i=1}^n \var\big(\epsilon_i B_{n,i}^g\big|\bs X)=\frac{4}{n^2}\sum_{i=1}^n \E(\epsilon_i^2) \big(B_{n,i}^g\big)^2=\frac{4}{n^2}\frac{3}{2p^2}\sum_{i=1}^n \big(B_{n,i}^g\big)^2\\
    &\leq \frac{6}{np^2}\max_{i\in\{1,\cdots,n\}}\big\{|B^g_{n,i}|^2\big\}\leq C\frac{h^4}{n}.
\end{align*}
For $a_n=h^2/\sqrt{n}$ and all $\delta>0$, an application of Chebyshev's inequality conditionally to $\bs X$ implies that
\begin{equation*}
    P\big(\big|B_{n}^{(g,\epsilon)}\big|>a_n\delta\big|\bs X\big)\leq C\frac{h^4}{n\delta^2a_n^2}\leq \frac{C}{\delta^2}.
\end{equation*}
Then, by the Law of Iterated Expectations, it follows that
\begin{equation*}
    P\big(\big|B_{n}^{(g,\epsilon)}\big|>a_n\delta\big)=\E\big(\E(I(|B_{n}^{(g,\epsilon)}\big|>a_n\delta)|\bs X)\big)\leq \frac{C}{\delta^2},
\end{equation*}
 which shows that $B_{n}^{(g,\epsilon)}=O_p(h^2/\sqrt{n})$.

 We now turn to the stochastic term $V_{n}^{(g,\epsilon)}$. Its conditional mean can be expressed  as
 \begin{align} \label{eqnormal}
    \E\big(V_{n}^{(g,\epsilon)}|\bs X\big)&= \frac{2}{n}\sum_{i,j=1}^n W_j(X_i)\E(\epsilon_i\epsilon_j)=\frac{2}{n}\sum_{i=1}^n W_i(X_i)\E(\epsilon_i^2)+\frac{2}{n}\sum_{i=1}^n \sum_{\substack{j=1\\j\neq i}}^n W_j(X_i)\underbrace{\E(\epsilon_i)\E(\epsilon_j)}_{=0}\nonumber\\
    &=\frac{2}{n}\E(\epsilon_1^2)\sum_{i=1}^n W_i(X_i).
 \end{align}
To compute the conditional variance of $V_{n}^{(g,\epsilon)}$ given $\bs X$, we need to investigate the expression
\begin{align*}
    \big(V_{n}^{(g,\epsilon)}\big)^2&= \frac{4}{n^2}\bigg(\sum_{i,j=1}^n W_j(X_i)\epsilon_i\epsilon_j\bigg)^2=\frac{4}{n^2}\sum_{(i,j,\ell,r)\in[n]^4}W_j(X_i)W_\ell(X_r)\epsilon_i\epsilon_j\epsilon_\ell\epsilon_r,
 \end{align*}
where $[n]\coloneqq\{1,\cdots,n\}$. As $\{\epsilon_i\}_{i=1}^n$ is a collection of independent centered random variables, it is sufficient to evaluate only the subset, say $T$, of $[n]^4$ consisted of all 4-tuples that have two pairs of entries with identical indices. The subset of indices $T\subset[n]^4$ can be partitioned into four  subsets, that is, $T=\bigcup_{i=1}^4 T_i$, where
\begin{align*}
    T_1&=\{(i,j,\ell,r)\in[n]^4: i=j,\ell=r,i\neq \ell\}, \quad T_2=\{(i,j,\ell,r)\in[n]^4: i=\ell,j=r,i\neq j\}\\
    T_3&=\{(i,j,\ell,r)\in[n]^4: i=r,j=\ell,i\neq j\}, \quad  T_4=\{(i,j,\ell,r)\in[n]^4: i=j=\ell=r\}.
\end{align*}
This allows us to rewrite
\begin{align*}
    \frac{n^2}{4}\big(V_{n}^{(g,\epsilon)}\big)^2&=\sum_{(i,j,\ell,r)\in T_1}W_i(X_i)W_\ell(X_\ell)\epsilon_i^2\epsilon_\ell^2
    +\sum_{(i,j,\ell,r)\in T_2}W_j(X_i)W_i(X_j)\epsilon_i^2\epsilon_j^2\\
    &\quad + \sum_{(i,j,\ell,r)\in T_3}W_j(X_i)^2\epsilon_i^2\epsilon_j^2+\sum_{(i,j,\ell,r)\in T_4}W_i(X_i)^2\epsilon_i^4.
\end{align*}
Consequently, it follows that
\begin{align}\label{eqnormal2}
    \frac{n^2}{4}\E\Big(\big(V_{n}^{(g,\epsilon)}\big)^2\big|\bs X\Big)&
    =\big[\E(\epsilon_1^2)\big]^2\sum_{i=1}^n \sum_{\substack{\ell=1\\\ell\neq i}}^n W_i(X_i)W_\ell(X_\ell)
    +\big[\E(\epsilon_1^2)\big]^2\sum_{i=1}^n \sum_{\substack{j=1\\j\neq i}}^nW_j(X_i)W_i(X_j)+\nonumber \\
    &\qquad + \big[\E(\epsilon_1^2)\big]^2\sum_{i=1}^n \sum_{\substack{j=1\\j\neq i}}^n W_j(X_i)^2
    +\E(\epsilon_1^4)\sum_{i=1}^nW_i(X_i)^2\nonumber\\
&=\big[\E(\epsilon_1^2)\big]^2\Bigg[\sum_{i,j=1}^n\big(W_j(X_i)^2+W_j(X_i)W_i(X_j)\big)+\bigg(\sum_{i=1}^n W_i(X_i)\bigg)^2\Bigg]+\nonumber\\
&\qquad + \Big(\E(\epsilon_1^4)-3\big[\E(\epsilon_1^2)\big]^2\Big)\sum_{i=1}^nW_i(X_i)^2,
\end{align}
where, for the second equality in \eqref{eqnormal2}, we have used the following three facts:
\begin{align*}
    &\sum_{i=1}^n \sum_{\substack{\ell=1\\\ell\neq i}}^n W_i(X_i)W_\ell(X_\ell)=\bigg(\sum_{i=1}^n W_i(X_i)\bigg)^2-\sum_{i=1}^n W_i(X_i)^2;\\
    &\sum_{i=1}^n \sum_{\substack{j=1\\j\neq i}}^nW_j(X_i)W_i(X_j)=\sum_{i,j=1}^n W_j(X_i)W_i(X_j)-\sum_{i=1}^n W_i(X_i)^2;
\end{align*}
and
\begin{equation*}
\sum_{i=1}^n \sum_{\substack{j=1\\j\neq i}}^n W_j(X_i)^2=-\sum_{i,j=1}^n W_j(X_i)^2-\sum_{i=1}^n W_i(X_i)^2.
\end{equation*}
Combining \eqref{eqnormal} and \eqref{eqnormal2}, we immediately have
\begin{align}\label{eqnormal3}
\var\Big(V_{n}^{(g,\epsilon)}\big|\bs X\Big)&= \E\Big(\big(V_{n}^{(g,\epsilon)}\big)^2\big|\bs X\Big)-\Big[\E\big(\big(V_{n}^{(g,\epsilon)}\big)\big|\bs X\big)\Big]^2\nonumber\\
&=\frac{4}{n^2}\big[\E(\epsilon_1^2)\big]^2\bigg(\sum_{i,j=1}^n\big[W_j(X_i)^2+W_j(X_i)W_i(X_j)\big]\bigg)+\nonumber\\
&\qquad + \frac{4}{n^2}\Big(\E(\epsilon_1^4)-3\big[\E(\epsilon_1^2)\big]^2\Big)\sum_{i=1}^nW_i(X_i)^2.
\end{align}
We proceed establishing the asymptotic orders of the mean and the variance, conditioned to $\bs X$. From \eqref{eqnormal},
\begin{equation}\label{eqnormal4}
    \E\big(V_{n}^{(g,\epsilon)}|\bs X\big)\leq \frac{3}{np^2}\sum_{i=1}^n |W_i(X_i)|\leq C\frac{n}{n(nh)}=O\bigg(\frac{1}{nh}\bigg),
\end{equation}
since $\displaystyle{\max_{i\in[n]}\Big\{\sup_{x\in\R}\big\{|W_i(x)|\big\}\Big\}\leq C/(nh)}$ \citep[Lemma 1.3 of][]{tsybakov}.
Moreover, from \eqref{eqnormal3},
\begin{align*}
\var\big(V_{n}^{(g,\epsilon)}\big|\bs X\big)&\leq\frac{9}{n^2p^4}\bigg[\sum_{i,j=1}^n\big(W_j(X_i)^2+|W_j(X_i)||W_i(X_j)|\big)\bigg]+ \frac{4}{n^2}\bigg[\frac{63}{4p^4}-3\frac{9}{4p^4}\bigg]\sum_{i=1}^nW_i(X_i)^2\\
&\leq \frac{C}{n^2(nh)^2}\sum_{i=1}^n\bigg[\sum_{j=1}^n \big|W_j(X_i)\big|\bigg]+\frac{Cn}{n^2(nh)^2}\leq C\bigg[\frac{1}{n^2h}+\frac{1}{n^3h^2}\bigg]=O\bigg(\frac{1}{n^2h}\bigg).
\end{align*}
Taking $b_n=1/(n\sqrt{h})$, for all $\delta>0$,
\begin{equation*}
    P\Big(\big|V_{n}^{(g,\epsilon)}-\E\big(V_{n}^{(g,\epsilon)}| \bs X\big)\big|>\delta b_n \Big|\bs X\Big)\leq\frac{\var\big(V_{n}^{(g,\epsilon)}\big|\bs X\big)}{(\delta b_n)^2}\leq \frac{C}{\delta^2},
\end{equation*}
by the conditional Chebyshev's inequality . This implies, by the Law of Iterated  Expectations, that
\begin{equation}\label{eqnormal5}
    V_{n}^{(g,\epsilon)}-\E\big(V_{n}^{(g,\epsilon)}| \bs X\big)=O_p\bigg(\frac{1}{n\sqrt{h}}\bigg).
\end{equation}
Using the results \eqref{eqnormal4} and \eqref{eqnormal5}, it follows that
\begin{align*}
    V_{n}^{(g,\epsilon)}&=\Big[V_{n}^{(g,\epsilon)}-\E\big(V_{n}^{(g,\epsilon)}|\bs X\big)\Big]+\E\big(V_{n}^{(g,\epsilon)}|\bs X\big)=O_p\bigg(\frac{1}{n\sqrt{h}}\bigg)+O\bigg(\frac{1}{nh}\bigg)=O_p\bigg(\frac{1}{nh}\bigg).
\end{align*}
Finally, we turn to the term $T_{2,n}$.  By the  Central Limit Theorem, it follows that
\begin{equation*}
\sqrt{n}\big(\tilde W-\theta\big)=\sqrt{n}\bigg(\frac1n\sum_{i=1}^n \epsilon_i^2-\E(\epsilon_1^2)\bigg)\overset{d}{\longrightarrow} N(0,\sigma^2).
\end{equation*}
where $\sigma^2\coloneqq\var(\epsilon_1^2)=27/(2p^4)$. Thus, from the results established above, it follows that
\begin{align*}
    \sqrt{n}(\hat W-\theta)&=\sqrt{n}(T_{1,n}+T_{2,n})=\sqrt{n}O_p\bigg(h^4+\frac{\ln(n)}{nh}+\frac{h^2}{\sqrt{n}}+\frac{1}{nh}\bigg)+\sqrt{n}T_{2,n}\\
    &=o_p(1)+\sqrt{n}T_{2,n}\overset{d}{\longrightarrow} N(0,\sigma^2)
\end{align*}
since $h\overset{a}{\approx} c_h n^{-c}$, for some $c\in(1/8,1/3)$, by assumption. Recall that $\theta=3/(2p^2)$. Upon considering the function $\ell_1(u)\coloneqq \sqrt{\frac3{2u}}$, whose first derivative is $\ell_1'(u)=-\sqrt{\frac{3}{8u^3}}$, the Delta method \citep[Theorem 3.1 in][]{vaart} yields
\begin{equation*}
\sqrt{n}(\hat p-p)=\sqrt{n}\big(\ell_1(\hat W)-\ell_1(\theta)\big)\overset{d}{\longrightarrow} N(0,\sigma^2[\ell_1'(\theta)]^2)=N(0,3p^2/2).
\end{equation*}
Since $\sqrt{n}(\hat p-p)$ converges to a Gaussian limit, it is bounded in probability \citep[theorem 2.4(i) of][]{vaart}, i.e., $\sqrt{n}(\hat p-p)=O_p(1)$, and so, $\hat p-p=O_p(1/\sqrt{n})$ as desired.
\qed

\proof{teo2.n}
 Theorem \ref{teo1} and the results obtained in the proof of Theorem 3.2 for $\hat p$ imply respectively that $\displaystyle{\max_{x\in[a,b]}}\big\{\big|\hat g(x)-g(x)\big|\big\}=o_p(1 )$ and $|\hat p-p|=o_p(1 )$, since $a_n=o(1)$ by hypothesis. Furthermore, for all $c>0$,
\begin{align*}
    P\bigg(\max_{x\in[a,b]}\Big\{\big|&\ln\big(\hat f(x)/a_n\big)-\ln\big(f(x)/a_n\big)\big|\Big\}>c\bigg)=\\
    &=P\bigg(\max_{x\in[a,b]}\Big\{\Big|\ln(\hat f(x))-\ln\Big(e^{\frac3{2p}-g(x)}\Big)\Big|\Big\}>c\bigg)\\
    &= P\bigg(\max_{x\in[a,b]}\bigg\{\bigg| \frac{3}{2\hat p}-\frac{3}{2p} -\hat g(x)+g(x)\bigg|\bigg\}>c\bigg)\\
    &\leq
    P\bigg(\bigg|\frac{3}{2\hat p}-\frac{3}{2p}\bigg|>\frac{c}{2} \bigg)
    +P\bigg(\max_{x\in[a,b]}\Big\{\Big| \hat g(x)-g(x)\Big|\Big\}>\frac{c}{2}\bigg)\coloneqq   B_{1,n}+B_{2,n}.
\end{align*}
It is clear that $B_{2,n}=o(1)$. Now since $\ell_2(u)\coloneqq  \frac3{2u}$ is a continuous function on $(0,\infty)$ and $\hat p>0$ almost surely, then the continuous mapping theorem \citep[Theorem 2.3(ii) of][]{vaart} implies that $B_{1,n}=o(1)$ as well. Hence, $\ln(\hat f(x)/a_n)-\ln(f(x)/a_n)=o_p(1)$ uniformly in $x\in[a,b]$. Let $\ell_3(u)\coloneqq e^u$. For all $x\in[a,b]$ and $c>0$, the continuity of $l_3(\cdot)$ implies that there is $c_2>0$ such that
\begin{align}\label{equa27}
P\Big(\big|\hat f(x)-f(x)\big|>ca_n\Big)&=P\bigg(\big|\ell_3\big(\ln\big(\hat f(x)/a_n\big)\big)-\ell_3\big(\ln\big(f(x)/a_n\big)\big)|>c\bigg)\nonumber\\
&\leq P\bigg(\big|\ln\big(\hat f(x)/a_n\big)-\ln\big(f(x)/a_n\big)|>c_2\bigg)+o(1)=o(1).
\end{align}
 Since \eqref{equa27} holds for all $x\in[a,b]$, it also holds for
\begin{equation*}
x_0\coloneqq \underset{x\in[a,b]}{\argmax}\big\{\big|\hat f(x)-f(x)\big|\big\}.
\end{equation*}
 Now, we proceed to analyze the asymptotic normality. Define $\hat D_{n,x}=3/(2\hat p)+\hat g(x)$ and $D_{x}=3/(2 p)+g(x)$. Firstly, assume that the bandwidth is chosen suboptimally, meaning that $h=o(n^{-1/5})$. Under this assumption, on one hand $\sqrt{nh}\big(\hat g(x)-g(x)-B(x)\big)=\sqrt{nh}\big(\hat g(x)-g(x)\big)+o(1)\overset{d}{\longrightarrow}N\big(0,V(x)\big)$ where the bias $B(x)$ and the variance $V(X)$ are defined in Theorem \ref{teo1}. Therefore, the bias decreases faster than $(nh)^{-1/2}$, becoming negligible as the sample size grows. On the other hand, $\sqrt{n}(\hat p-p)\overset{d}{\longrightarrow}N(0,3p^2/2)$ which implies that  $\sqrt{n}\big(\frac3{2\hat p}-\frac3{2p}\big)\overset{d}{\longrightarrow}N\big(0,\frac{2p^6}3\big)$, and hence $\frac3{2\hat p}-\frac3{2p}=O_p(n^{-1/2})$. Combining these two results, we find that
\begin{align*}
    \sqrt{nh}(\hat D_{n,x}-D_{x})&=\sqrt{nh}\bigg(\frac{3}{2\hat p}-\frac{3}{2 p}\bigg)-\sqrt{nh}\big(\hat g(x)-g(x)\big)\\
    &=o_p(1)-\sqrt{nh}\big(\hat g(x)-g(x)\big)\overset{d}{\longrightarrow}N(0,V(x)).
\end{align*}
The Delta method gives
\begin{align*}
    \sqrt{nh}\big(\hat f(x)-f(x)\big)&=\sqrt{nh}\big(\ell_3(\hat D_{n,x})-\ell_3(D_{x})\big)\overset{d}{\longrightarrow}N\big(0,V(x)f(x)^2\big).
\end{align*}
When the bandwidth is chosen optimally, that is, $h\overset{a}{\approx}c_h n^{-1/5}$ for some constant $c_h>0$, the bias is no longer negligible, but converges to $\mu_x\coloneqq\displaystyle{\lim_{n\to\infty}}\sqrt{nh}B(x)=\sqrt{c_h}g''(x)\int_\R u^2 K(u)du/2$.
In this case, upon combining the Slutsky's theorem and the Delta method, yields
\begin{align*}
    \sqrt{nh}(\hat f(x)-f(x))&=\sqrt{nh}(\ell_3(\hat D_{n,x})-\ell_3(D_{x}))\overset{d}{\longrightarrow}f(x)N\big(\mu_x,V(x)\big).
\end{align*}
and the proof is complete.\qed

\proof{teo3}
Following the proof of Theorem 3.3, we
define $\hat W=\sum_{i=1}^n\hat \epsilon_i^2/n$, $\tilde W=\sum_{i=1}^n \epsilon_i^2/n$, and $\theta=3/(2p^2)$. Then
\begin{equation*}
    \hat W-\theta= (\hat W-\tilde W)+(\tilde W-\theta)\coloneqq T_{1,n}+T_{2,n}.
\end{equation*}
We first show that $T_{1,n}=o_p(n^{-1/2})$. Write this term as
\begin{equation*}
    T_{1,n}=\frac{1}{n}\sum_{i=1}^n\big[g(X_i)-\hat g(X_i)\big]^2+\frac{2}{n}\sum_{i=1}^n \big[g(X_i)-\hat g(X_i)\big]\epsilon_i\coloneqq P_{1,n}+P_{2,n}.
\end{equation*}
After using the Cauchy-Schwarz inequality, we decompose the upper bound for $P_{1,n}$ into interior and boundary parts as follows:
\begin{align*}
    P_{1,n}&=\frac{1}{n}\sum_{i=1}^n\bigg[g_0-\hat g_0 +\sum_{j=1}^m g(X_{i,j})-\hat g(X_{i,j})\bigg]^2\\
    &\leq (m+1)(g_0-\hat g_0)^2+\frac{m+1}{n}\sum_{i=1}^n\sum_{j=1}^m\big[ g(X_{i,j})-\hat g(X_{i,j})\big]^2\\
    &\leq 2m(g_0-\hat g_0)^2+\frac{2m}{n}\sum_{j=1}^m\sum_{i\in I_{n,j}}\big[ g(X_{i,j})-\hat g(X_{i,j})\big]^2+\frac{2m}{n}\sum_{j=1}^m\sum_{i\in I_{n,j}^c}\big[ g(X_{i,j})-\hat g(X_{i,j})\big]^2\\
    &\coloneqq 2m\Big(P_{1,n}^{(0)}+P_{1,n}^{(I)}+P_{1,n}^{(B)}\Big),
\end{align*}
where $I_{n,j}\coloneqq\{i\in[n]:X_{i,j}\in i_n\}$, $I_{n,j}^c\coloneqq\{i\in[n]:X_{i,j}\in b_n\}$, $i_{ n,j}\coloneqq [h_j,1-h_j]$, $b_{ n,j}\coloneqq [0,h_j)\cup (1-h_j,1]$ and $[n]=\{1,\cdots,n\}$, for all $n\in\N$. Denote the cardinalities of $I_{ n,j}$ and $I_{ n,j}^c$ by $n_I=\sum_{i=1}^n I(X_{i,j}\in i_{ n,j})$ and $n_{I^c}=\sum_{i=1}^n I(X_{i,j}\in b_{ n,j})$, respectively. Then
\begin{align*}
    \E\bigg(\frac{n_{I^c}}{n}\bigg)=P(X_{i,j}\in b_{n,j}), \quad \mbox{ and }\quad \var\bigg(\frac{n_{I^c}}{n}\bigg)=\frac{P(X_{i,j}\in b_{n,j})-[P(X_{i,j}\in b_{n,j})]^2}{n}\leq \frac{1}{4n}.
\end{align*}
Hence, Chebyshev's inequality implies that, for all $\delta>0$,
\begin{equation*}
    P\bigg(\frac{n_{I^c}}{n}-P(X_{i,j}\in b_{n,j})>\frac{\delta}{\sqrt{n}}\bigg)\leq \frac{n\var(n_{I^c}/n)}{\delta^2} \leq\frac{1}{4\delta^2},
\end{equation*}
and thus, $n_{I^c}/n=P(X_{i,j}\in b_{n,j}) +O_p(n^{-1/2})$. Since $n_I=n-n_{I^c}$, we also have  $n_{I}/n=1-P(X_{i,j}\in b_{n,j}) +O_p(n^{-1/2})$. By assumption, the design density $f_{\bs X}$ is bounded above on $[0,1]^m$, hence each marginal $f_{X_j}$ is bounded above on $[0,1]$. Therefore,
\begin{equation*}
    P(X_{i,j}\in b_{n,j})=\int_{0}^{h_j} f_{X_j}(x_j)dx_j+\int_{1-h_j}^1 f_{X_j}(x_j)dx_j\leq 2Ch_j.
\end{equation*}
Since $h_j\overset{a}{\approx}c_{h,j}n^{-1/5}$, we obtain $n_{I^c}/n=O(h)+O_p(n^{-1/2})=O_p(n^{-1/5})$ and $n_I/n=1+O(h)+O_p(n^{-1/2})=1+O_p(n^{-1/5})$. Using  \eqref{equasb26} and \eqref{equasb26c} in Theorem \ref{teo2.2}, we conclude that
\begin{align*}
    &P^{(I)}_{1,n}\leq \frac{n_I}{n}\Big(\max_{x_j\in[h_j,1-h_j]} \Big\{\big|\hat g_j(x_j)-g_j(x_j)\big|\Big\}\Big)^2=(1+O_p(n^{-1/5}))O_p(n^{-4/5})=O_p(n^{-4/5}),\\
    &P^{(B)}_{1,n}\leq \frac{n_{I^c}}{n}\Big(\max_{x_j\in[0,1]} \Big\{\big|\hat g_j(x_j)-g_j(x_j)\big|\Big\}\Big)^2=O_p(n^{-1/5})O_p(n^{-2/5})=O_p(n^{-3/5}).
\end{align*}
Moreover, since $g_0-\hat g_0=g_0-1/n\sum_{i=1}^n Z_i$, we have that $\E(g_0-\hat g_0)=0$ and $\var(g_0-\hat g_0)=\var(Z_1)/n\leq \E(Z_1^2)/n<\infty$ by Assumption \textbf{B3}. By the Central Limit Theorem, $\hat g_0=g_0+O_p(n^{-1/2})$, and hence $P^{(0)}_{1,n}=O_p(n^{-1})$. Therefore, $P_{1,n}=O_p(n^{-1})+O_p(n^{-4/5})+O_p(n^{-3/5})=O_p(n^{-3/5})=o_p(n^{-1/2})$.

Next, we turn to the term $P_{2,n}$. Using the decomposition \eqref{eqdec1}-\eqref{eqdec2}, as in the proof of Theorem \ref{teo2.2}, we write $\hat g_j(x_j)=\hat g_j^B(x_j)+\hat g_j^S(x_j)$, for all $j\in[m]$, and then
\begin{align*}
    P_{2,n}&=\frac{2}{n}\sum_{i=1}^n \big[g(X_i)-\hat g^B(X_i)\big]\epsilon_i+\frac{2}{n}\sum_{i=1}^n \hat g^S(X_i)\epsilon_i\\
    =&\ (g_0-\hat g^B_0)\frac{2}{n}\sum_{i=1}^n\epsilon_i+\frac{2}{n}\sum_{i=1}^n\sum_{j=1}^m \big[g_j(X_{i,j})-\hat g_j^B(X_{i,j})\big]\epsilon_i+\hat g^S_0\frac{2}{n}\sum_{i=1}^n\epsilon_i+\frac{2}{n}\sum_{i=1}^n\sum_{j=1}^m\hat g_j^S(X_{i,j})\epsilon_i\\
    \coloneqq&\ P_{2,n}^{(B,0)}+P_{2,n}^{(B,1)}+P_{2,n}^{(S,0)}+P_{2,n}^{(S,1)}.
\end{align*}
The term $P_{2,n}^{(B,0)}$ is the product of two independent random variables
\begin{equation*}
    P_{2,n}^{(B,0)}=(g_0-\hat g^B_0)\frac{2}{n}\sum_{i=1}^n\epsilon_i=\bigg[\sum_{j=1}^m \frac{1}{n}\sum_{\ell=1}^n g_j(X_{\ell,j}) \bigg] \bigg[\frac{2}{n}\sum_{i=1}^n\epsilon_i\bigg].
\end{equation*}
By the identification assumption, $\E\big(g_j(X_{\ell,j})\big)=0$, $\forall j\in[m],\ell\in[n]$, and by Assumption \textbf{B4} each $g_j$ and $f_{X_j}$, is continuous on $[0,1]$, $\forall j\in[m]$. Hence, the Weierstrass approximation theorem implies that
\begin{align*}
    \E\big([g_j(X_{\ell,j})]^2\big) = \int_0^1[g_j(x_j)]^2f_{X_j}(x_j)dx_j\leq C<\infty, \quad \forall j\in[m],\ell\in[n],
\end{align*}
and thus $g_0-\hat g^B_0=O_p(n^{-1/2})$ by the Central Limit Theorem. Analogously, we can show that $2/n\sum_{i=1}^n\epsilon_i=O_p(n^{-1/2})$. Therefore, $P_{2,n}^{(B,0)}=O_p(n^{-1})=o_p(n^{-1/2})$.

To cope with the term $P_{2,n}^{(B,1)}$, consider the uniform rates \eqref{eqdec4} and \eqref{eqdec5}, which state that $\displaystyle{\max_{x_j\in[h_j,1-h_j]}}\big\{|\hat g_j(x_j)^B-g_j(x_j)|\big\}=O_p(n^{-2/5})$ and $\displaystyle{\max_{x_j\in[0,1]}}\big\{|\hat g_j(x_j)^B-g_j(x_j)|\big\}=O_p(n^{-1/5})$, respectively. Distinguishing interior and boundary design points as before, we obtain
\begin{align*}
    B_n&\coloneqq\frac{1}{n}\sum_{i=1}^n\bigg[\sum_{j=1}^m \big(g_j(X_{i,j})-\hat g_j^B(X_{i,j})\big)\bigg]^2\\
    &\leq m\frac{n_I}{n}\Big(\max_{x_j\in[h_j,1-h_j]} \Big\{\big|g_j(x_j)-\hat g_j^B(x_j)\big|\Big\}\Big)^2+m\frac{n_{I^c}}{n}\Big(\max_{x_j\in[0,1]} \Big\{\big|g_j(x_j)-\hat g_j^B(x_j)\big|\Big\}\Big)^2\\
&=O_p\big(n^{-4/5}\big)+O_p\big(n^{-3/5}\big)=O_p\big(n^{-3/5}\big).
\end{align*}
This shows that $\forall \delta>0:\exists C_\delta>0: P(B_n\geq C_\delta n^{-3/5})\leq \delta$ for all sufficiently large $n$. Decompose
\begin{equation*}
    P_{2,n}^{B,1}= P_{2,n}^{B,1}I(B_n\geq C_\delta n^{-3/5})+P_{2,n}^{B,1}I(B_n<C_\delta n^{-3/5})\coloneqq \underline P_{2,n}^{(B,1)}+\overline P_{2,n}^{(B,1)}.
\end{equation*}
Since $B_n$ is $\sigma(\bs X)$-measurable, it follows that, conditionally on $\bs X$,
\begin{equation*}
    \E\Big(\overline P_{2,n}^{(B,1)}\big|\bs X\Big)=I\big(B_n<C_\delta n^{-3/5}\big)\frac{2}{n}\sum_{i=1}^n\sum_{j=1}^m \big(g_j(X_{i,j})-\hat g_j^B(X_{i,j})\big)\E(\epsilon_i)=0,
\end{equation*}
and
\begin{align*}
    \var\Big(\overline P_{2,n}^{(B,1)}\big|\bs X\Big)&=I\big(B_n<C_\delta n^{-3/5}\big)\var\Big(P_{2,n}^{(B,1)}\big|\bs X\Big)\\
    &=I\big(B_n<C_\delta n^{-3/5}\big)\frac{4}{n^2}\sum_{i=1}^n\bigg[\sum_{j=1}^m \big(g_j(X_{i,j})-\hat g_j^B(X_{i,j})\big)\bigg]^2\var(\epsilon_i)\\
    &\leq \frac{C}{n}I\big(B_n<C_\delta n^{-3/5}\big)B_n<Cn^{-8/5}.
\end{align*}
By the conditional Chebyshev's inequality and the Law of Iterated Expectations,
\begin{equation*}
    P\Big(\overline P_{2,n}^{(B,1)}>\overline{C}n^{-1/2}\Big)\leq C n^{-3/5}.
\end{equation*}
for any $\overline{C}>0$. On the other hand,
\begin{equation*}
    P\Big(\underline P_{2,n}^{(B,1)}>\overline{C}n^{-1/2}\Big)=P\Big(P_{2,n}^{(B,1)}>\overline{C}n^{-1/2},B_n\geq C_\delta n^{-3/5}\Big)\leq P\Big(B_n\geq C_\delta n^{-3/5}\Big) \leq \delta,
\end{equation*}
for all sufficiently large $n$. Therefore,
by the law of total probability,
\begin{align*}
    P\Big( P_{2,n}^{(B,1)}>\overline{C}n^{-1/2}\Big)&=P\Big( \overline P_{2,n}^{(B,1)}>\overline{C}n^{-1/2}\Big) +P\Big( \underline P_{2,n}^{(B,1)}>\overline{C}n^{-1/2}\Big)\\
    &\leq C n^{-3/5}+\delta=o(1)+\delta,
\end{align*}
and since $\delta,\overline{C}>0$ are arbitrary constants, it follows that $P_{2,n}^{(B,1)}=o_p(n^{-1/2})$.

For $P_{2,n}^{(S,0)}$, we have
\begin{equation*}
    \E\Big(P_{2,n}^{(S,0)}\Big)=\E\bigg(\frac{2}{n^2}\sum_{i,\ell=1}^n \epsilon_i\epsilon_\ell\bigg )=\frac{2}{n^2}\sum_{i=1}^n \E\big(\epsilon_i^2\big) =\frac{3}{np^2}=O\big(n^{-1}\big),
\end{equation*}
and, along the lines of \eqref{eqnormal3} in the proof of Theorem \ref{teo2}, we can check that
\begin{align*}
    \var\Big(P_{2,n}^{(S,0)}\Big)=\frac{4}{n^4}\big[\E(\epsilon_1^2)\big]^2\big\{2n^2\big\}+\frac{4}{n^4}\Big(\E(\epsilon_1^4)-3\big[\E(\epsilon_1^2)\big]^2\Big)n=O\big(n^{-2}\big)+O\big(n^{-3}\big)=O\big(n^{-2}\big).
\end{align*}
By Chebyshev's inequality, $P_{2,n}^{(S,0)}=O(n^{-1})+O_p(n^{-1})=O_p(n^{-1})=o_p(n^{-1/2})$.

Finally, for $P_{2,n}^{(S,1)}$, using the uniform expansion \eqref{eqdec4}, as in the proof of Theorem \ref{teo2.2}, namely, $\hat g_j(x_j)^S=\tilde g_{j}^S(x_j)+\frac{1}{n}\sum_{i=1}^n r_{ij}(x_j)\epsilon_i+o_p(n^{-1/2})$, and upon defining $D_n(x_j)\coloneqq \hat g_j^S(x_j)-\tilde g_j^S(x_j)-1/n\sum_{i=1}^n r_{i,j}(x_j)\epsilon_{i}$, we  write
\begin{align*}
   P_{2,n}^{(S,1)}&=\sum_{j=1}^m\bigg(\frac{2}{n}\sum_{i=1}^n\tilde g_{j}^S(X_{i,j})\epsilon_i+\frac{2}{n^2}\sum_{i,\ell=1}^nr_{\ell j}(X_{i,j})\epsilon_\ell\epsilon_i+\frac{2}{n}\sum_{i=1}^nD_n(X_{i,j})\epsilon_i\bigg)\\
   &\coloneqq\sum_{j=1}^m\Big( V_{n,j}^{(NW)}+V_{n,j}^{(r)}+V_{n,j}^{(D)} \Big).
\end{align*}
Since $\var(|\epsilon_1|)\leq\E(|\epsilon_1|^2)=3/(2p^2)<\infty$, the Central Limit Theorem implies that
\begin{equation*}
    \big|V_{n,j}^{(D)}\big|\leq 2\sup_{x_j\in[0,1]}\big\{|D_n(x_j)|\big\}\frac{1}{n}\sum_{i=1}^n|\epsilon_i|=o_p(n^{-1/2})O_p(1)=o_p(n^{-1/2}).
\end{equation*}
Since each function $r_{\ell j}$ is uniformly bounded, it follows that
\begin{equation*}
    \E\Big(V_{n,j}^{(r)}\big|\bs X \Big)=\frac{2}{n^2}\E(\epsilon_1^2)\sum_{i=1}^n r_{ij}(X_{i,j})\leq \frac{C}{n}.
\end{equation*}
Arguing as in \eqref{eqnormal3}, in the proof of Theorem \ref{teo2}, we obtain that
\begin{align*}\notag
\var\Big(V_{n,j}^{(r)}\big|\bs X\Big)
&=\frac{4}{n^4}\big[\E(\epsilon_1^2)\big]^2\Bigg(\sum_{i,\ell=1}^n\big[r_{\ell j}(X_{i,j})^2+r_{\ell j}(X_{i,j})r_{i j}(X_{\ell,j})\big]\Bigg)\\
&\quad + \frac{4}{n^4}\Big(\E(\epsilon_1^4)-3\big[\E(\epsilon_1^2)\big]^2\Big)\sum_{i=1}^nr_{i j}(X_{i,j})^2\leq C\bigg(\frac{1}{n^2}+\frac{1}{n^3}\bigg)\leq \frac{C}{n^2}.
\end{align*}
Chebyshev's inequality gives $V_{n,j}^{(r)}=O(n^{-1})+O_p(n^{-1})=o_p(n^{-1/2})$, whereas the term $V_{n,j}^{(NW)}$ can be bounded exactly as $V_{n}^{(g,\epsilon)}$ in the proof of Theorem \ref{teo2}, with the weights $W_i$ replaced by the Nadaraya-Watson weights $W_i(x_j)=K_{h_j}(x_j, X_{i,j})/\sum_{\ell=1}^n K_{h_j}(x_j, X_{\ell,j})$. We can then show that $ V_{n,j}^{(NW)}=O_p(n^{-4/5})=o_p(n^{-1/2})$. Upon combining the results obtained so far, we conclude that $T_{1,n}=o_p(n^{-1/2})$. From the proof of Theorem \ref{teo2}, $\sqrt{n}T_{2,n}\overset{d}{\to} N\big(0,27/(2p^4)\big)$, and by the delta method we conclude that $\sqrt{n} (\hat p-p)\overset{d}{\to} N(0,3p^2/2)$. Hence, $\hat p-p=O_p(n^{-1/2})$ as claimed.
The proof for the uniform convergence  and asymptotic normality of $\hat f(x)-f(x)$ follows by the same arguments used to prove Theorem \ref{teo2.n}, and thus is omitted. \qed

 \noindent \textbf{Proof of Lemma \ref{lem1}}  Let $G_i\coloneqq-\ln(R_i)\sim Gamma(3/2,1/p)$. Its cumulative distribution function $G$ can be expanded through a Maclaurin series as
 \begin{align*}
     F_G(x)&=\int_0^{px}\sqrt{\frac{u}{\pi}}e^{-u}du=\int_0^{px}\sqrt{\frac{u}{\pi}}\sum_{k=0}^\infty \frac{(-u)^k}{k!}du=\sum_{k=0}^\infty \frac{(-1)^k}{k!\sqrt{\pi}}\int_0^{px}u^{k+1/2}du\\
     &=\sum_{k=0}^\infty \frac{(-1)^k}{k!\sqrt{\pi}}\frac{(px)^{k+3/2}}{k+3/2}=\sqrt{\frac{(px)^3}{\pi}}\sum_{k=0}^\infty \frac{(px)^k}{k!(k+3/2)}=\sqrt{\frac{(px)^3}{\pi}}\big[1+O(x)\big].
 \end{align*}
 as $x$ approches 0 from the right. Then, since $1-x\leq e^{-x}$ for all $x\geq 0$, we have that
 \begin{align*}
     P\Big(\min_{1\leq i\leq n}\{G_i\}>&C_1 n^{-2/3}\Big)=\big(1-F_G(C_1 n^{-2/3})\big)^n\leq \exp\big\{-nF_G(C_1 n^{-2/3})\big\}\\
     &=\exp\bigg\{-\sqrt{\frac{(pC_1)^3}{\pi}}\big[1+O(n^{-2/3})\big]\bigg\}\leq \exp\bigg\{-\sqrt{\frac{(pC_1)^3}{\pi}}[1-\delta_1]\bigg\}
 \end{align*}
 for all $C_1>0$, some arbitrarily small $\delta_1>0$ and all sufficiently large $n$. As $C_1>0$ can be taken arbitrarily large, it follows that $\displaystyle{\min_{1\leq i\leq n}}\{G_i\}=O_p(n^{-2/3})$.   \qed

\proof{teo36}
Since $\hat\epsilon_i=\hat\epsilon_i\pm \epsilon_i=\epsilon_i+g(X_i)-\hat g(X_i)$, it follows that
\begin{equation*}
    \min_{1\leq i\leq n}\{\epsilon_i\}-\max_{1\leq i\leq n}\big\{\big|g(X_i)-\hat g(X_i)\big|\big\}\leq \min_{1\leq i\leq n}\{\hat\epsilon_i\}\leq \min_{1\leq i\leq n}\{\epsilon_i\}+\max_{1\leq i\leq n}\big\{\big|g(X_i)-\hat g(X_i)\big|\big\},
\end{equation*}
and, hence,
\begin{align*}
    \Big| \min_{1\leq i\leq n}\{\hat \epsilon_i\}- \min_{1\leq i\leq n}\{\epsilon_i\}\Big|\leq  \max_{1\leq i\leq n}\big\{\big|g(X_i)-\hat g(X_i)\big|\big\}\leq \sup_{x\in[a,b]}\big|g(x)-\hat g(x)\big|=O_p\bigg(h^2+\sqrt{\frac{\ln(n)}{nh}}\bigg).
\end{align*}
Adding and subtracting $\min_{1\leq i\leq n} \{\epsilon_i\}$, Lemma \ref{lem1} yields
\begin{align}\label{eqe}
    \min_{1\leq i\leq n}\{\hat \epsilon_i\}&
    =O_p\bigg(h^2+\sqrt{\frac{\ln(n)}{nh}}\bigg)-\frac{3}{2p}+\min_{1\leq i\leq n}\big\{-\ln(R_i)\big\}\nonumber\\
    &=-\frac{3}{2p}+O_p\bigg(n^{-2/3}+h^2+\sqrt{\frac{\ln(n)}{nh}}\bigg)=-\frac{3}{2p}+O_p\bigg(h^2+\sqrt{\frac{\ln(n)}{nh}}\bigg),
\end{align}
as $n^{-2/3}=o\big(\sqrt{\ln(n)/(nh)}\big)$.
Define $\ell_3(u)\coloneqq -3/(2u)$. From the Mean Value Theorem, we obtain that
\begin{equation}\label{eqe2}
    \hat p_m-p=\ell_3\Big(\min_{1\leq i\leq n}\{\hat \epsilon_i\}\Big)-\ell_3\bigg(-\frac{3}{2p}\bigg)=\ell_3'(\xi_n)\bigg(\min_{1\leq i\leq n}\{\hat \epsilon_i\}+\frac{3}{2p}\bigg),
\end{equation}
where $\xi_n\coloneqq -3/(2p)+\tau_n\Big[\displaystyle{\min_{1\leq i\leq n}}\hat \epsilon_i+3/(2p)\Big]$ for some $\tau_n\in(0,1)$. Thus, using \eqref{eqe},
\begin{align*}
    \bigg|\xi_n+\frac{3}{2p}\bigg|\leq \bigg|\min_{1\leq i\leq n}\{\hat \epsilon_i\}+\frac{3}{2p}\bigg|=O_p\bigg(h^2+\sqrt{\frac{\ln(n)}{nh}}\bigg)=o_p(1).
\end{align*}
Then, the application of the continuous mapping theorem implies that $\ell_3'(\xi_n)$  converges in probability to $\ell_3'(-3/(2p))=2p^2/3$, and thus,
\begin{equation}\label{eqe3}
\ell_3'(\xi_n)=2p^2/3+o_p(1)=O(1)+o_p(1)=O_p(1).
\end{equation}
From equation \eqref{eqe2}, we use results \eqref{eqe} and \eqref{eqe3} to obtain that
\begin{equation*}
    \hat p_m-p=O_p(1)O_p\bigg(h^2+\sqrt{\frac{\ln(n)}{nh}}\bigg)=O_p\bigg(h^2+\sqrt{\frac{\ln(n)}{nh}}\bigg).
\end{equation*}
The proof for the multivariate case follows analogously, except that we have to replace the rate of convergence of the local linear estimator, $h^2+\sqrt{\ln(n)/(nh)}$, with the smooth backfitting rate, $n^{-1/5}$.

Finally, to show that $P(\hat p_m>0)\to 1$ as $n\to\infty$, it suffices to notice that $\hat p_m>0$ if, and only if, $\displaystyle{\min_{1\leq i\leq n}\{\hat \epsilon_i\}<0}$ and that
in both settings we have
$\displaystyle{\min_{1\leq i\leq n}}\{\hat \epsilon_i\}+3/(2p)=o_p(1)$, and the result follows straightforwardly.\qed
%

\proof{teo37}
We start by showing that
\begin{equation*}
    |\hat Q_{n,\hat\epsilon}-Q_{n,\epsilon}|\leq |\hat Q_{n,\hat\epsilon}-\hat Q_{n,\epsilon}|+|\hat Q_{n,\epsilon}-Q_{n,\epsilon}|=O_p\big(b_n\big),
\end{equation*}
where $b_n\coloneqq \max\{a_n,n^{-1/2}\alpha_n^{1/6}\}$ and $a_n\coloneqq h^2+\sqrt{\ln(n)/(nh)}$. Since $\hat \epsilon_i=\epsilon_i+g(X_i)-\hat g(X_i)$ and $\displaystyle{\max_{{x\in[0,1]}}}\big\{|g(x)-\hat g(x)|\big\}\leq Ca_n$ (Theorem \ref{teo1}), we have for all $i\in\{1,\cdots,n\}$,
\begin{equation*}
    \epsilon_i-Ca_n\leq \hat \epsilon_i\leq \epsilon_i+Ca_n\Longrightarrow\epsilon_{(i)}-Ca_n\leq \hat \epsilon_{(i)}\leq \epsilon_{(i)}+Ca_n\Longrightarrow \hat Q_{n,\epsilon}-Ca_n\leq \hat Q_{n,\hat\epsilon}\leq \hat Q_{n,\epsilon}+Ca_n,
\end{equation*}
or, equivalently, $|\hat Q_{n,\hat \epsilon}-\hat Q_{n,\epsilon}|\leq Ca_n$.
From the Bahadur' representation \cite[][Theorem 5.11]{shao}, we have
\begin{equation}\label{eqquant2}
    \hat Q_{n,\epsilon}=Q_{n,\epsilon}+\frac{F_\epsilon(Q_{n,\epsilon})-\hat F_\epsilon(Q_{n,\epsilon})}{f_\epsilon(Q_{n,\epsilon})}+o_p(n^{-1/2}),
\end{equation}
where $\hat F_\epsilon(x)\coloneqq 1/n\sum_{i=1}^nI(\epsilon_i\leq x)$, denotes the empirical distribution of $\{\epsilon_i\}_{i=1}^n$. Since $F_\epsilon$ is continous on $\R$, $\E(\hat F_\epsilon(Q_{n,\epsilon}))=F_\epsilon(Q_{n,\epsilon})=\alpha_n$ and $\var(\hat F_\epsilon(Q_{n,\epsilon}))=\alpha_n(1-\alpha_n)/n$. By Chebyshev's inequality,
\begin{equation}\label{eqquant3}
    F_\epsilon(Q_{n,\epsilon})-\hat F_\epsilon(Q_{n,\epsilon})=O_p\big(\sqrt{\alpha_n/n}\big).
\end{equation}
Moreover,
\begin{equation}\label{eqquant4}
    f_\epsilon\big(Q_{n,\epsilon}\big)=f_{-\ln(R)}\big(Q_{n,\epsilon}+3/(2p)\big)=f_{-\ln(R)}\big(\gamma_n/p\big)
\end{equation}
Since $\gamma^{-1}(3/2,\cdot)$ is strictly increasing on its domain $\big(0,\sqrt{\pi}/2\big)$ and  $\alpha_n=o(1)$, we have that $\gamma_n=\gamma^{-1}(3/2,\alpha_n\sqrt{\pi}/2)=o(1)$. Also,
$\gamma\big(3/2,\gamma^{-1}(3/2,\alpha_n\sqrt{\pi}/2)\big)=\alpha_n\sqrt{\pi}/2$ by definition, and \cite[see 8.354.1 in][]{grad}
\begin{equation*}
    \gamma(3/2,x)=\sum_{k=0}^\infty \frac{(-1)^k x^{3/2+k}}{k!(3/2+k)}=\frac{2x^{3/2}}{3}\big(1+o(1)\big), \ \text{ as }x\to 0.
\end{equation*}
Thus, as $n\to\infty$,
\begin{equation}\label{eqquant5}
    \frac{2\gamma_n^{3/2}}{3}\big(1+o(1)\big)=\frac{\alpha_n\sqrt{\pi}}{2}\implies \gamma_n\overset{a}{\approx}\bigg(\frac{3\alpha_n\sqrt{\pi}}{4}\bigg)^{2/3}.
\end{equation}
By Maclaurin's expansion and \eqref{eqquant5},
\begin{equation}\label{eqquant6}
    e^{-\gamma_n}=1-\gamma_n+o(\gamma_n)=1+O\big(\alpha_n^{2/3}\big).
\end{equation}
Therefore, from \eqref{eqquant4}-\eqref{eqquant6},
\begin{equation}\label{eqquant7}
f_\epsilon(Q_{n,\epsilon})=2p\sqrt{\frac{\gamma_n}{\pi}}e^{-\gamma_n}=\sqrt{\gamma_n}\Big(1+O\big(\alpha_n^{2/3}\big)\Big)\overset{a}{\approx}\bigg(\frac{3\alpha_n\sqrt{\pi}}{4}\bigg)^{1/3}
\end{equation}
Combining equations \eqref{eqquant2}, \eqref{eqquant3} and \eqref{eqquant7} we obtain
\begin{equation*}
    |\hat Q_{n,\epsilon}-Q_{n,\epsilon}|=O_p\big(\alpha_n^{1/6}n^{-1/2}\big)+o_p\big(n^{-1/2}\big)=O_p\big(\alpha_n^{1/6}n^{-1/2}\big).
\end{equation*}
Hence,
\begin{equation*}
     |\hat Q_{n,\hat\epsilon}-Q_{n,\epsilon}|=O_p(a_n)+O_p\big(\alpha_n^{1/6}n^{-1/2}\big)=O_p(b_n).
\end{equation*}
We proceed showing that $\hat p_\alpha-p=O_p(b_n)$. As $\gamma_n=o(1)$, $Q_{n,\epsilon}=(\gamma_n-3/2)/p=-3/(2p)+o(1)$. Thus, $|Q_{n,\epsilon}+3/(2p)|<\delta$ for all $\delta>0$ and all $n$ sufficiently large. By taking $\delta=3/(4p)$, the triangle inequality gives, for $n$ large enough,
\begin{equation*}
    |Q_{n,\epsilon}|\geq \frac{3}{2p}-\bigg |Q_{n,\epsilon}+\frac{3}{2p}\bigg |>\frac{3}{2p}-\frac{3}{4p}=\frac{3}{4p}>0.
\end{equation*}
Hence, if $|\hat Q_{n,\hat \epsilon}-Q_{n,\epsilon}|<3/(8p)$, then
\begin{equation*}
    |\hat Q_{n,\hat \epsilon}|\geq |Q_{n,\epsilon}|-|\hat Q_{n,\hat \epsilon}-Q_{n,\epsilon}|>\frac{3}{4p}-\frac{3}{8p}=\frac{3}{8p}>0,
\end{equation*}
for all $n$ large enough. Consequently, by the monotonicity of the probability measure,
\begin{align*}
    P\bigg(|\hat Q_{n,\hat \epsilon}|\geq \frac{3}{8p}\bigg)\geq P\bigg(|\hat Q_{n,\hat \epsilon}-Q_{n,\epsilon}|< \frac{3}{8p}\bigg)=1-P\bigg(|\hat Q_{n,\hat \epsilon}-Q_{n,\epsilon}|\geq \frac{3}{8p}\bigg)=1+o(1),
\end{align*}
for all $n$ sufficiently large, since $|\hat Q_{n,\hat\epsilon}-Q_{n,\epsilon}|=o_p(1)$. This implies that $P\big(|\hat Q_{n,\hat \epsilon}|< 3/(8p)\big)=o(1)$.

Define $\ell_5(u)\coloneqq (\gamma_n-3/2)/u$. If $|\hat Q_{n,\hat \epsilon}|\geq 3/(8p)$ and $n$ is large enough so that $|Q_{n,\epsilon}|>3/(4p)$, then the Fundamental Theorem of Calculus gives
\begin{align*}
    \hat p_\alpha-p&=\ell_5\big(\hat Q_{n,\hat \epsilon}\big)-\ell_5\big( Q_{n, \epsilon}\big)=\int_{Q_{n, \epsilon}}^{\hat Q_{n,\hat \epsilon}}\ell_5'(u)du\\
    &\leq \sup_{u:|u|\geq 3/(8p)}\big\{\big|\ell_5'(u)\big|\big\}\big|\hat Q_{n,\hat \epsilon}- Q_{n, \epsilon}\big|\leq \underbrace{|\gamma_n-3/2|\bigg(\frac{8p}{3}\bigg)^{2}}_{=O(1)}O_p(b_n)=O_p(b_n),
\end{align*}
since $\gamma_n-3/2\to-3/2$. We have shown that $\forall\delta_1>0:\exists M>0:$
\begin{equation*}
    P(|\hat p_\alpha-p|\geq b_n M,|\hat Q_{n,\hat \epsilon}|\geq 3/(8p))\leq \delta_1,
\end{equation*}
for all $n$ sufficiently large. Therefore, using the Total Law of Probability, the monotonicity of the probability measure and the fact that $P\big(|\hat Q_{n,\hat \epsilon}|< 3/(8p)\big)=o(1)$, it follows that
\begin{align*}
    P(|\hat p_\alpha-p|\geq b_n M)&\leq P(|\hat p_\alpha-p|\geq b_n M,|\hat Q_{n,\hat \epsilon}|\geq 3/(8p))+P(|\hat Q_{n,\hat \epsilon}|< 3/(8p))\leq 2\delta_1
\end{align*}
for all $n$ large enough, as desired. Finally, we show that $b_n=O(a_n)$. By hypothesis, $\alpha_n\to 0$ and there exist $c_h>0$ and $v\in(0,1)$ such that $h\to c_h n^{-v}$. Thus, for $n$ large enough, $h<2c_h n^{-v}$, $\sqrt{\ln(n)/(2c_h)}>1$ and $\alpha_n<1$. Hence, for $n$ large enough,
\begin{equation*}
    a_n>\sqrt{\frac{\ln(n)}{nh}}>n^{-(1-v)/2}\sqrt{\frac{\ln(n)}{2c_h}}>n^{-1/2}>\alpha_n^{1/6}n^{-1/2},
\end{equation*}
showing that $b_n=\max\{a_n,\alpha_n^{1/6}n^{-1/2}\}=a_n$ as soon as $n$ is sufficiently large.

Similar arguments can be used for the multivariate case, except that we have to consider the convergence rate $n^{-1/5}$ associated with the smooth backfitting estimator. To show that $\hat p_\alpha$ lies in the parameter space $(0,\infty)$ with probability tending to one, note that $|\hat p_\alpha-p|<p/2\implies \hat p_\alpha>p/2\implies 0<\hat p_\alpha$. Hence, by the monotonicity of the probability measure,
\begin{equation*}
    P(\hat p_\alpha>0)\geq P(\hat p_\alpha>p/2)\geq P(|\hat p_\alpha-p|<p/2)\to 1,
\end{equation*}
since $|\hat p_\alpha-p|=o_p(1)$.
\section*{Appendix B. The Matsuoka's distribution.}\label{mat}
Let $p>0$ be a parameter and consider the probability density function given by \eqref{eq1}, which is reproduced here for completeness:
\begin{equation*}
f_p(x)\coloneqq  2\sqrt{\frac{-p^3 \ln(x)}{\pi}} x^{p-1} I(x\in(0,1)).
\end{equation*}
This function is non-negative and it is easy to show that it integrates 1 over $\R$ \citep[see formula 4.269.3 in][]{grad}. A two-parameter probability density related to \eqref{eq1}  was first considered in \cite{consul}, as a particular transformation of the generalized gamma distribution introduced in \cite{stacy}, to the unit interval $(0,1)$. \cite{consul} called it the log-gamma distribution and derived a few simple properties of \eqref{eq1}, such as moments and the distribution of some specific functions of independent log-gamma distributed variates. The main results in the paper are related to certain likelihood ratio tests  for which the authors were able to derive asymptotic results. Later \cite{Grassia}, independently, introduced the same distribution, derived its moments and presented some real data applications. In the applications, the author claimed that the introduced distribution presented similar properties as the beta distribution, which was later discussed in detail in \cite{griff},  where \cite{Grassia}'s distributions was referred to as Grassia 1 and 2. Related distributions were also introduced with the name log-gamma distribution, but considering different methods. \cite{hogg} called the log-gamma distribution the exponential of a gamma distribution, while \cite{hell} considered the one obtained by taking the logarithm transformation of a gamma distribution. The distribution given by \eqref{eq1} is also a particular case of a very broad and general class of distributions introduced in \cite{ufpe}, called the Unit Gamma-G class. However, \cite{ufpe} does not elaborate upon the distribution \eqref{eq1}.

It is often the case that only few general results can be derived analytically for very general classes of distributions, such as the Unit Gamma-G. On the other hand, the focus on \cite{consul,Grassia, griff} was elsewhere and, despite its appearance in a few works, to the best of our knowledge there is very little work done after \cite{consul} directly related to \eqref{eq1}. Given the gap in the literature, in this section we shall provide some results regarding \eqref{eq1}. These properties play a central role in the asymptotic theory derived in Section \ref{at}. Since the name log-gamma was attributed to at least 3 different distributions  we shall provide a name for \eqref{eq1} and say that a random variable $X$ taking values in $(0,1)$ follows a Matsuoka distribution with parameter $p>0$, denoted $X\sim \mathrm{M}(p)$, if its density is given by \eqref{eq1}.

\subsubsection*{General properties}
\begin{figure}[ht!]
\centering
\includegraphics[width=0.5\textwidth]{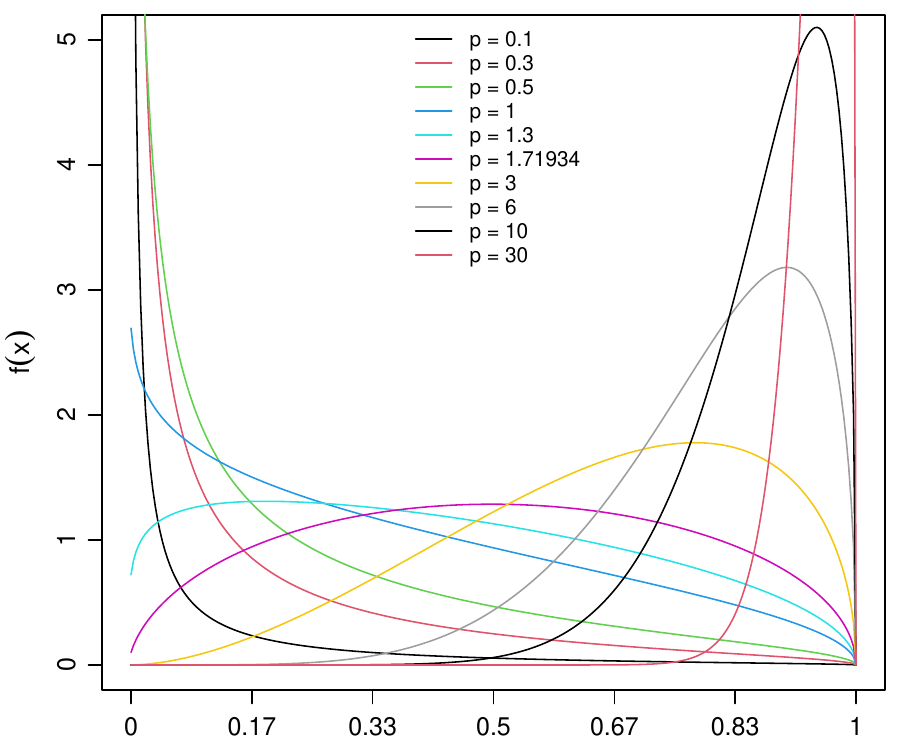}
\caption{ Plot of the density function for various parameters $p$.}\vskip.3cm\label{fig:density_mv}
\end{figure}

Figure \ref{fig:density_mv}  illustrates the different shapes that the density function $f_p$ can assume depending on the values of the parameter $p$. The graphs suggest that small values for $p$ are associated with right-skewed functions, while large values for $p$ are associated with left-skewed functions. To be more precise, for $p\leq 1$ the density of a M$(p)$ distribution assumes a J-shaped pattern, while for $p>1$ the density is unimodal with mode located at $x=e^{-\frac1{2(p-1)}}$ with value $\sqrt{\frac{2p^3e^{-1}}{\pi(p-1)}}$.
Indeed, considering $\ln(f_p(x))$, it is easy to see that
\begin{equation*}
\frac{\partial\ln(f_p(x))}{\partial x}=\frac{2(p - 1) \ln(x) + 1}{2x\ln(x)}<0,\  \forall x\in(0,1)\qquad \Longleftrightarrow \qquad p\leq 1,
\end{equation*}
so that \eqref{eq1} assumes a J-shape pattern if, and only if, $p\leq 1$. For $p>1$,  we have
\begin{equation*}
\frac{\partial\ln(f_p(x))}{\partial x}=0 \qquad  \Longleftrightarrow\qquad  x=e^{-\frac1{2(p-1)}}.
\end{equation*}
Now, the second derivative is given by
\begin{equation*}
\frac{\partial^2\ln(f_p(x))}{\partial x^2}= -\frac{2 (p - 1) \ln(x)^2 + \ln(x) + 1}{2 x^2 \ln(x)^2}<0,
\end{equation*}
since, for $p>1$, the polynomial $2 (p - 1) y^2 + y + 1$ is an upward parabola with complex roots, hence, strictly positive. We conclude that \eqref{eq1} is unimodal for $p>1$ with $x=e^{-\frac1{2(p-1)}}$ as mode. In particular, \eqref{eq1} is never symmetric about $1/2$. To see this, observe that, to be symmetric about $1/2$, the mode must be located at $x=1/2$. Hence the only option is $p_0=1+\frac{1}{2\ln(2)}$. Also, to be symmetric, \eqref{eq1} must satisfy
\begin{equation*}
f_{p_0}(x)=f_{p_0}(1-x), \ \ \forall x\in(0,1/2) \qquad \Longleftrightarrow\qquad\bigg(\frac{x}{1-x}\bigg)^{\frac1{\ln(2)}} =\frac{\ln(1-x)}{\ln(x)},\ \ \forall x\in(0,1/2),
\end{equation*}
which is, of course, absurd, since the last equality does not hold for any $x\in(0,1/2)$.

To calculate the cumulative distribution function related to \eqref{eq1},  $F_p(x)$, anti-differentiation based on formula 8.356.4 in \cite{grad} yields
\begin{align}\label{eq2}
F_p(x)=\frac{2}{\sqrt{\pi}}\Gamma\bigg(\frac32,-p\ln(x)\bigg)I(0<x<1)+I(x\geq1),
\end{align}
for $x\in\R$, where $\Gamma(k,t)=\int_t^\infty z^{k-1}e^{-z}dz$ for all $k>0$, is the upper incomplete gamma function \citep[see section 8.35 in][]{grad}.
From \eqref{eq2}, it is easy to calculate the associated quantile function, which is given by $F^{-1}_p(0)=0$, $F^{-1}_p(1)=1$ and
\begin{equation}\label{qu}
F^{-1}_p(q)=\exp\bigg\{-\frac1p\Gamma^{-1}\bigg(\frac32,\frac{q\sqrt{\pi}}2\bigg)\bigg\},
\end{equation}
for $q\in(0,1)$, where $\Gamma^{-1}(k,x)$ denotes the inverse of the upper incomplete gamma function, satisfying $\Gamma\big(k,\Gamma^{-1}(k,x)\big)=\Gamma^{-1}\big(k,\Gamma(k,x)\big)=x$, for all $x\in\R$ and $k>0$. Of course, the upper incomplete gamma function and its inverse can only be computed numerically.

The moments associated with $X\sim\mathrm{M}(p)$ are easily calculated. Upon changing variables to $z=\sqrt{-\ln(t)}\ \Rightarrow\ dt=-2ze^{-(t+p)z^2}dz$ and applying integration by parts, it follows that
\begin{align}\label{eqee}
\E(X^k)&=2\sqrt{\frac{p^3}{\pi}} \int_0^1 \sqrt{-\ln(t)}t^{p+k-1}dt=\frac4{\sqrt{\pi}}\bigg(\frac{p}{p+k}\bigg)^{\frac32}\int_0^\infty z^2e^{-(p+k)z^2}dz=\bigg(\frac{p}{p+k}\bigg)^{\frac32},
\end{align}
for all $k>-p$. In particular, from \eqref{eqee} we have,
\begin{align*}
&\E(X)=\bigg(\frac{p}{p+1}\bigg)^{\frac32},\qquad\var(X)=\bigg(\frac{p}{p+2}\bigg)^{\frac32}-\bigg(\frac{p}{p+1}\bigg)^{3}, \\ &\E\big([X-\E(X)]^3\big)=\bigg(\frac p{p+3}\bigg)^{\frac32}-\frac{3p^3}{\big[(p+1)(p+2)\big]^{\frac32}}+2\bigg(\frac p{p+1}\bigg)^{\frac92}.
\end{align*}
 The existence of all moments of order $k\in\N$ allows the calculation of the moment-generating function as
\begin{align*}
  M_X(t)=\sum_{n=0}^\infty \E(X^n) \frac{t^n}{n!}=p^{3/2}\sum_{n=0}^\infty \frac{t^n}{n!(p+n)^{3/2}}.
\end{align*}
Observe that the last expression is bounded by $e^t$, for all $t\in\R$ and $p>0$. Calculating the moment-generating function using $\E(e^{tX})$ and \eqref{eq1} is unfeasible.
\subsubsection*{Exponential family and point estimation}
It is easy to see that Matsuoka's distribution is a member of the 1-parameter regular exponential family,
in the form $f_p(x)=h(x)\exp\{\eta t(x)-a(\eta)\}$,
with canonical parameter $\eta=p$, $a(\eta)= -\frac32\ln(\eta)$, $h(x)=-\frac{2\ln(x)}{x\sqrt{\pi}}I(0\leq x\leq 1)$,  and natural complete sufficient statistics given by $T(X)=\ln(X)$ \citep[][section 1.6]{bickel}. It is easy to show that if $X\sim\mathrm M(p)$, then $-\ln(X)\sim \mathrm{Gamma}\big(\frac32,\frac1p\big)$ (in shape/scale parametrization).

Let $X_1,\cdots,X_n$ be a sample from $X\sim\mathrm{M}(p)$. The log-likelihood for $p$ is given by
\begin{equation}\label{ml}
\ell(p)=n\ln\biggl(\frac2{\sqrt{\pi}}\bigg)+\frac{3n}2\ln(p)+\frac12\sum_{i=1}^n \ln\bigl(-\ln(X_i)\big) + (p-1)\sum_{i=1}^n\ln(X_i).
\end{equation}
Upon deriving \eqref{ml} with respect to $p$ and equating to 0, the maximum likelihood estimator (MLE) is obtained in closed-form and is given by $\hat p=-\frac{3n}{2\sum_{i=1}^n\ln(X_i)}$.
Now since  $-\ln(X)\sim \mathrm{Gamma}\big(\frac32,\frac1p\big)$, it follows that $-\sum_{i=1}^n\ln(X_i)\sim \mathrm{Gamma}\big(\frac{3n}2,\frac1p\big)$, so that $-\big[\sum_{i=1}^n\ln(X_i)\big]^{-1}\sim \mbox{Inv-Gamma}\big(\frac{3n}2,p\big)$. Hence, we conclude that the MLE is biased with $\E(\hat p)=\big(\frac{3n}{3n-2}\big)p$. Since the MLE is a function of the complete sufficient statistics $T(\bs X)=\sum_{i=1}^n\ln(X_i)$, it follows by the Lehmann-Scheff\'e theorem that the biased corrected estimator
\[\hat p^\ast= -\frac{3n-2}{2\sum_{i=1}^n\ln(X_i)}\]
is the uniformly minimum variance unbiased estimator for $p$, with $\var(\hat p^\ast)=\frac{3p^2}{3n-4}$.

Since the production frontier model defined in \eqref{equa2} is in log scale, the estimation procedure in Section \ref{pf} relies on the centered Gamma random variable $Y\coloneqq-\ln(X)-\E(-\ln(X))$, where $-\ln(X)\sim \mathrm{Gamma}\big(\frac32,\frac1p\big)$. Then $\E(Y)=0$ and $\var(Y)=\E(Y^2)=3/(2p^2)$. Let $Y_1,\cdots,Y_n$ be a sample from $Y$. A method of moments estimator of $p$ is obtained by equating the second sample and theoretical moments as follows:
\begin{equation}\label{eqqq}
    \E(Y^2)=\frac{3}{2p^2}=\frac{1}{n}\sum_{i=1}^n Y_i^2\implies \hat p_{mm}=\sqrt{\frac{3n}{2\sum_{i=1}^n Y_i^2}}.
\end{equation}
The method of moments estimator in \eqref{eqqq} is used in the parametric part  of our semiparametric estimation procedure (Section \ref{sec.a2}).

Due to page restrictions, further properties of the Matsuoka's distribution are presented in the supplementary material that accompanies the paper. To be more precise, we derive the skewness and kurtosis, order statistics, incomplete moments, mean deviation, entropy and a closed-formula for the stress-strength reliability.

\includepdf[pages=1-11]{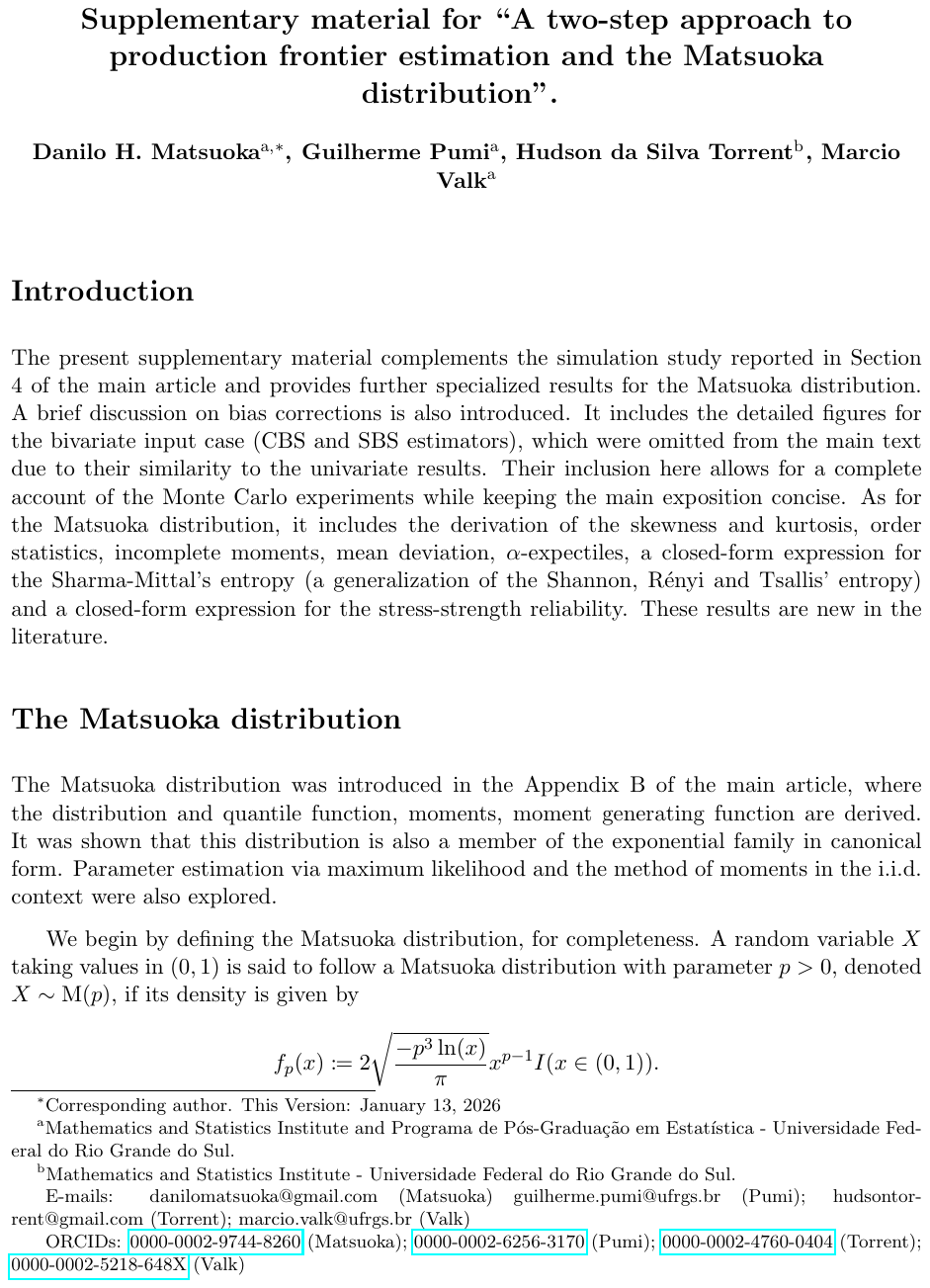}



\end{document}